\documentclass[aps,pra,showpacs,twocolumn,superscriptaddress]{revtex4-2}
\usepackage{amsmath}
\usepackage{amsfonts}
\usepackage{amssymb}
\usepackage{graphicx}
\usepackage{bbm}
\usepackage{natbib}
\usepackage{braket}
\usepackage[%
bookmarksnumbered,
bookmarksopen,
colorlinks,
citecolor=blue,
linkcolor=blue,
urlcolor=blue,
]{hyperref}
\makeatletter

\begin{document}

\title{The transfer of nonlocality between two- and three-qubit dissipative systems with counter-rotating-wave terms}

\author{Zi-Yu Xiong}
\affiliation{School of Physics and Electronic Science, Guizhou Normal University, Guiyang 550025, China}
\author{Yong-Jun Xiao}
\affiliation{School of Physics and Electronic Science, Guizhou Normal University, Guiyang 550025, China}
\author{Ye-Qi Zhang}
\affiliation{Department of Mathematics and Physics, North China Electric Power University, Beijing 102206, China}
\author{Qi-Liang He}\email{heliang005@163.com}
\affiliation{School of Physics and Electronic Science, Guizhou Normal University, Guiyang 550025, China}

\begin{abstract}
  We investigate the effect of counter-rotating-wave terms on nonlocality and entanglement for three qubits coupled with a common bath for strong and ultrastrong coupling regimes beyond the traditional treatment of Born-Markovian, perturbative and rotating wave approximations by employing the numerical hierarchical equations of motion approach.
  Our findings are as follows: (\romannumeral1) In the strong coupling regime, the counter-rotating terms accelerate the decay of genuine three-party correlations, and the obvious sudden birth of BN is found; (\romannumeral2) In the ultrastrong coupling regime, we observe a novel phenomenon where nonlocality is consistently transferred between a three-qubit and its subsystem. Besides, the inclusion of counter-rotating wave terms obviously enhances genuine tripartite nonlocality; and (\romannumeral3) These counter-rotating terms cannot effectively generate genuine three-party correlations in zero-excitation cases, which differs from previous studies involving only two qubits.
\end{abstract}

\maketitle

\section{Introduction}\label{sec:sec1}
Bell nonlocality (BN), as the fundamental concept of quantum mechanics, allows the tensorial non-separable bipartite states to correlate with each other even in the space-like interval, which can arise from quantum entanglement~\cite{PhysRev.47.777,bell1964einstein,RevModPhys.86.419}. 
In previous studies, quantum entanglement and BN for two parties have been extensively studied experimentally~\cite{shalm2021device,PhysRevLett.126.050503,10.1063/5.0179566,PhysRevLett.67.661,PhysRevLett.70.1895} and theoretically~\cite{PhysRevA.101.042112,PhysRevA.97.022111,PhysRevLett.125.200501,PhysRevLett.125.180505}, demonstrating their advantages over classical physics. 
However, the quantiﬁcation of multipartite entanglement and nonlocality becomes much more complicated~\cite{PhysRevA.62.062314}. To address this, a new important concept referred to as ``genuine'' has been introduced for multipartite systems~\cite{PhysRevLett.127.040403}, which includes genuine tripartite entanglement (GTE)~\cite{PhysRevResearch.4.023059,PhysRevA.110.032420,PhysRevA.83.062325} and nonlocality (GTN)~\cite{PhysRevD.35.3066}. 
As experiments progress, results indicate that both the maximum entangled three-qubit W~\cite{PhysRevA.81.052334} state and the GHZ state~\cite{PhysRevLett.102.250404} exhibit GTN and GTE correlations, revealing that these correlations have benefits for processing quantum information tasks~\cite{PhysRevLett.129.060401,kang2016fast,wei2015preparation,neeley2010generation,PhysRevA.96.022121,arnon2018practical,jozsa2003role,PRXQuantum.2.020304}.
For example, GTN demonstrates a unique application value since it can work against a conspiring (cheating) sub-group of parties in applying quantum communication complexity~\cite{PhysRevA.96.022121,arnon2018practical}, and GTE is crucial for quantum algorithms to achieve an exponential speed-up over classical computation~\cite{jozsa2003role,PRXQuantum.2.020304}.
 
However, GTN and GTE may be destroyed rapidly due to the unavoidable coupling between the microcosmic quantum system and its surrounding environment, which is the main challenge in quantum information processing~\cite{breuer2002theory}. The study of quantum open system dynamics has gained significant attention in recent decades due to its potential to simulate various physical processes~\cite{RevModPhys.89.015001}. 
Traditionally, the dynamics of most exactly solvable open quantum system models involve various approximations, such as Born-Markovian, perturbative, and rotating wave approximations (RWA)~\cite{nielsen2010quantum,weiss2012quantum}. These approximations are valid when the coupling between the system and the environment is weak and the time scale of the environment is significantly shorter than that of the system~\cite{breuer2002theory}. 
Unfortunately, the above approximations do not accurately predict real physical processes in the strong coupling regime because they fail to describe the dynamics of the environment completely~\cite{PhysRevA.96.032125,PhysRevA.85.062323,PhysRevA.97.052309}. 
For example, many studies have examined the dynamics of two- or three-qubit quantum correlations in non-Markovian environments with significant memory effects~\cite{PhysRevLett.100.090503,Nourmandipour,PhysRevA.79.032310}. In these cases, the correlation in the quantum system may revive due to the information flowing back from the environment, which is a typical characteristic of a non-Markovian quantum process.
 However, the dynamics of those studies are based on the RWA, which neglects the counter-rotating-wave terms; other studies that include the effects of counter-rotating wave terms rely on perturbative approximations~\cite{PhysRevA.82.022119,PhysRevLett.101.200404,PhysRevA.81.042116}, which are not applicable in cases of strong coupling regime.
 Naturally, we have an interesting question: What effects do the counter-rotating wave terms have on entanglement and nonlocality of three-qubit systems in a strong coupling regime without the approximations above?
 Here, we investigate the time evolution of three-qubit systems interacting with a common bath using numerical hierarchical equations of motion (HEOM)~\cite{doi:10.1143/JPSJ.58.101,PhysRevE.75.031107,10.1063/5.0011599}. 
 The HEOM is a precise method to describe the non-Markovian dynamics of a system interacting with one or more baths at finite temperatures without the approximations above, which is widely applied in the study of quantum dissipative systems~\cite{PhysRevA.96.032125,PhysRevA.85.062323,PhysRevA.97.052309,PhysRevLett.104.250401,PhysRevA.95.042132,PhysRevA.94.062116}. 
 
 In this paper, we utilize the HEOM scheme to investigate a model consisting of three noninteracting qubits that couple to a common bosonic bath. This model has been solved precisely for two qubits under the RWA~\cite{PhysRevA.79.032310}. Here, we extend the analysis from the two-qubit case to the three-qubit scenario and obtain an exact solution. By comparing the results with and without the RWA, we analyze the impact of the counter-rotating wave terms on the dynamic evolution of the entanglement and nonlocality of the reduced three-qubit system.
 In the strong coupling regime, we observe that the counter-rotating wave terms notably increase the decoherence rate of the reduced qubit system and diminish the sudden birth strength of GTE and GTN that arises from the memory effect of the bath.
 Besides, the sudden birth of BN is found for the two-qubit subsystem during the evolution, which indicates that the nonlocality is transferred from the whole three-qubit system to its subsystem.
Interestingly, we find that the counter-rotating wave terms can enhance the sudden birth of GTN but do not have the same effect on GTE for the ultrastrong coupling regime.
 Additionally, a novel phenomenon is observed regarding the continuous transfer of nonlocality between the three-qubit system and its subsystem by varying the coupling strength between the qubits and the bath.
 We also explore the case where the initial state has no excitation.
The results indicate that while entanglement can still be generated within two-qubit subsystems, neither GTE nor GTN can be effectively induced, which suggests that the virtual excitations produced by the counter-rotating wave terms cannot significantly generate genuine three-party correlations.

 This paper is organized as follows. In Sec.~\ref{sec:sec2}, we first introduce the methods used in this paper to quantify the entanglement and nonlocality in two-qubit and three-qubit systems.
  Next, we introduce the model studied in this paper and outline the steps for obtaining the time evolution under the non-RWA and RWA approaches.
 In Sec.~\ref{sec:sec3}, we investigate the influence of the counter-rotating-wave terms on the entanglement and BN in both strong and ultrastrong coupling regimes. The main conclusions of this paper are presented in Sec.~\ref{sec:sec4}.
 In the Appendix.~\ref{sec:secapp1} and~\ref{sec:secapp2}, we outline the steps for using the HEOM method to obtain the dynamics of the reduced qubit system without applying the RWA and provide the exact dynamic equation when the RWA is applied.

\section{Preliminaries}\label{sec:sec2}
\subsection{Entanglement and Bell nonlocality}
For a two-qubit system represented by the density matrix $\rho_{ab}$, there are several schemes for measuring the entanglement between 
qubits $a$ and $b$, including negativity~\cite{PhysRevA.65.032314}, discord~\cite{PhysRevLett.88.017901}, and 
concurrence~\cite{PhysRevLett.80.2245}. In this paper, we use concurrence to measure entanglement because it is a simple but reliable method widely employed in relevant research.
In quantum correlation, entanglement is considered a more fundamental type of correlation because it serves as the source of BN. While BN arises from entanglement, it is important to note that entanglement does not necessarily imply BN~\cite{RevModPhys.86.419,wu2022genuine}. To assess whether a two-qubit system exhibits the nonlocal correlation, we can apply the CHSH inequality~\cite{PhysRevLett.23.880}. The local correlated two-qubit state $\rho_{ab}$ must satisfy the CHSH inequality, which can be expressed as
\begin{equation}
  \operatorname{tr}(\rho_{ab} S) \leq 2,
\end{equation}
where $S$ is the CHSH operator, and the inequality can be calculated for any two-qubit density matrix using the following expression~\cite{PhysRevA.101.042112}:
\begin{align}
  \mathcal{N}(\rho_{ab}) & = \max _{S} \operatorname{tr}(\rho_{ab} S),\notag\\
  &=2 \sqrt{\lambda_{1}+\lambda_{2}},
  \end{align}
where $\mathcal{N}(\rho_{ab})$ denotes the maximal value of the CHSH expression for an arbitrary two-qubit state $\rho_{ab}$. if $\mathcal{N}(\rho_{ab}) > 2 $, the two-qubit state must be nonlocal. Besides,
$\lambda_{1}$ and $\lambda_{2}$ are the spectral decomposition of the matrix $T^{T} T=\sum_{k=1}^{3} \lambda_{k}\left|\mu_{k}\right\rangle\left\langle\mu_{k}\right|$ in nonincreasing order.
The $3 \times 3$ matrix $T$ is constructed using the Pauli operators given by the vector $\vec{\sigma} = (\sigma_{1}, \sigma_{2}, \sigma_{3})$, where the entries of the matrix are defined as $T_{jk} = \operatorname{tr}(\sigma_{j} \otimes \sigma_{k} \rho)$.

However, the definition of entanglement becomes more complex when considering a three-qubit system.
In recent decades, many GTE measures have been proposed~\cite{PhysRevResearch.4.023059,PhysRevA.110.032420,PhysRevA.83.062325,PhysRevLett.127.040403,Guo_2022,PhysRevA.61.052306}, but only a few ways to effectively calculate the GTE for general three-qubit states, primarily due to the challenges involved in identifying the optimal decompositions of mixed three-qubit states.
Thus, we choose the method of $\pi$-tangle to quantify the GTE since this method provides an explicit analytical expression for arbitrary three-qubit states and can be defined in the following manner~\cite{PhysRevA.75.062308,torres2019entanglement}:
\begin{align}
  \pi_{abc}=\frac{1}{3}\left(\pi_{a}+\pi_{b}+\pi_{c}\right),
\end{align}
where, $\pi$-tangle is the average of $\pi_a$, $\pi_b$ and $\pi_c$.
\begin{equation}\label{Eq:tangle}
\begin{array}{l}
  \pi_{a}=N_{a(bc)}^{2}-N_{ab}^{2}-N_{ac}^{2}, \\
  \pi_{b}=N_{b(ac)}^{2}-N_{ba}^{2}-N_{bc}^{2}, \\
  \pi_{c}=N_{c(ab)}^{2}-N_{ca}^{2}-N_{cb}^{2},
\end{array}
\end{equation}
where $N_{a(bc)}=\Vert \rho_{abc}^{T_{a}}\Vert-1$ and $N_{ab}=\Vert\rho_{ab}^{T_{a}}\Vert-1$ are the negativity for a two- and 
three-qubit state, respectively. $T_a$ denotes the partial transpose of $\rho_{abc}$ or $\rho_{ab}$, and $\Vert M \Vert$ is the trace norm for a
matrix M. The other symbols have the same definition like $N_{a(bc)}$ and $N_{ab}$.

As for GTN, it can be quantified by using the Svetlichny inequality~\cite{PhysRevD.35.3066}
\begin{equation}
  \operatorname{tr}(S \rho_{abc}) \leq 4,
\end{equation}
where the maximal value violating the Svetlichny inequality of the tripartite state $\rho_{abc}$ can be denoted as
\begin{align}
\mathcal{N}(\rho_{abc}) \equiv \max _{S} \operatorname{tr}(S \rho_{abc}).
\end{align}
If $\mathcal{N}(\rho_{abc})>4$, then the three-qubit system must exhibit GTN.
However, there is no specific analytic expression to calculate $\mathcal{N}(\rho_{abc})$ for generalized three-qubit state
due to computational complexity since here we use the ``exact'' numerical method to quantify the three-qubit states whether violating the Svetlichny inequality~\cite{xiong2024quantum}.
\subsection{The Model}
In this subsection, we derive the dynamics of the reduced density matrix for three non-interacting qubits in a common bath, considering both cases with and without RWA.
The Hamiltonian of the total system without RWA can be described by
\begin{equation}\label{Eq:model}
  \hat{H} = \omega_{0} \sum_{i=1}^{3}\left( \hat{\sigma}_{+}^{(i)}  \hat{\sigma}_{-}^{(i)}\right)+\sum_{k} \omega_{k} \hat{a}_{k}^{\dagger} \hat{a}_{k}+ \hat{V} \otimes \sum_{k} g_{k}\left( \hat{a}_{k}^{\dagger}+ \hat{a}_{k}\right),
\end{equation}
where $\hat{\sigma}_{ \pm}^{(j)}$ is the spin-flip operators, $\omega_0$ and $\omega_k$ denote the frequencies of qubit and bath,
$\hat{a}_{k}^{\dagger}$ and $\hat{a}_{k}$ are the creation and annihilation operators of the bath. Besides, $\hat{V}=\sum_{i=1}^{3} \alpha_i \hat{\sigma}_x^i$ denotes the qubits system coupled
to the bath, containing the counter-rotating-wave terms. The coupling strength of the $i$th qubit with the bath is denoted by $\alpha_i$.

In this paper, we examine the structured bath as the electromagnetic field within a lossy cavity, where the qubit-cavity coupling spectrum takes on a Lorentzian form due to the broadening of the fundamental mode~\cite{PhysRevA.79.032310}
\begin{equation}\label{Eq:spectrum}
  J(\omega)=\frac{1}{\pi} \frac{\lambda \gamma}{\left(\omega-\omega_{0}\right)^{2}+\gamma^{2}},
\end{equation}
here $\lambda$ is proportional to the vacuum Rabi frequency and can approximately interpreted as the system-bath coupling
strength, $\gamma$ is the width of the distribution and the quantity $\frac{1}{\gamma}$ is the lifetime of the mode.
Besides, the ratio $R=\frac{\alpha_t \sqrt{\lambda}}{\gamma}$ denotes the boundary between Markovian regimes and non-Markovian
regimes,where $\alpha_t$ is the collective coupling strength defined in Appendix.~\ref{sec:secapp2}. When $ R \gg 1 $, it indicates that the qubits and the bath are strongly coupled, which implies that the memory effects of the bath must be considered, and the dynamics of this open system are non-Markovian.
The time evolution of the reduced qubit system described by Eq.~(\ref{Eq:model}) can be analyzed using the HEOM method, with further details provided in Appendix~\ref{sec:secapp1}.

In the following, we will provide the exact solution to the above model under the RWA. Suppose the initial state of the system is 
\begin{equation}
  |\psi(0)\rangle=\ket{W}\otimes  |0\rangle_{R},
\end{equation}
where $\ket{W}=\frac{1}{\sqrt{3}}[\ket{egg}+\ket{geg}+\ket{gge}]$ and the time evolution is described by
\begin{align}
  |\psi(t)\rangle &  = c_{1}(t) |e, g,g\rangle|\mathbf{0}\rangle_{R}+c_{2}(t)  |g, e,g\rangle|\mathbf{0}\rangle_{R} \notag \\
  &+c_{3}(t)  |g, g,e\rangle|\mathbf{0}\rangle_{R} +\sum_{k} c_{k}(t) |g, g,g\rangle\left|1_{k}\right\rangle_R,
  \end{align}
  where $\left|1_{k}\right\rangle_R$ denotes the one excitation state in the $k$th mode. By tracing over the bath degrees of freedom, 
  the reduced three-qubit density matrix $\rho_{abc}$ takes the form
  \begin{equation}\label{Eq:RWA}
    \rho_{abc}= \left(\begin{array}{cccccccc}
      0 & 0 & 0 & 0 & 0 & 0 & 0 & 0 \\
      0 & 0 & 0 & 0 & 0 & 0 & 0 & 0 \\
      0 & 0 & 0 & 0 & 0 & 0 & 0 & 0 \\
      0 & 0 & 0 & \left|c_{1}\right|^{2} & 0 & c_{1} c^*_{2} & c_{1} c^*_{3} & 0 \\
      0 & 0 & 0 & 0 & 0 & 0 & 0 & 0 \\
      0 & 0 & 0 & c_{2} c^*_{1} & 0 & \left|c_{2}\right|^{2} & c_{2} c_{3}^* & 0 \\
      0 & 0 & 0 & c_{3} c^*_{1} & 0 & c_{3} c^*_{2} & \left|c_{3}\right|^{2} & 0 \\
      0 & 0 & 0 & 0 & 0 & 0 & 0 & \left|c_{k}\right|^{2}
      \end{array}\right).
    \end{equation}
    Then, following procedures shown Appendix.~\ref{sec:secapp2}, the amplitudes $c_i$ are obtained
    \begin{equation}
      \begin{aligned}
        c_{1}(t) & = \left[r_{2}^{2}+r_{3}^{2}+r_{1}^{2} \mathcal{E}(t)\right] c_{1}(0)-r_{1} r_{2}[1-\mathcal{E}(t)] c_{2}(0)\\ &-r_{1} r_{3}[1-\mathcal{E}(t)] c_{3}(0), \\
        c_{2}(t) & =\left[r_{1}^{2}+r^2_3+r_{2}^{2} \mathcal{E}(t)\right] c_{2}(0) -r_{1} r_{2}[1-\mathcal{E}(t)] c_{1}(0) \\ &-r_{2} r_{3}[1-\mathcal{E}(t)] c_{3}(0), \\
        c_{3}(t) & =  \left[r_{1}^{2}+r_{2}^{2}+r_{3}^{2} \mathcal{E}(t)\right] c_{3}(0)-r_{1} r_{3}[1-\mathcal{E}(t)] c_{1}(0)\\ & -r_{2} r_{3}[1-\mathcal{E}(t)] c_{2}(0).
        \end{aligned}
      \end{equation}

\section{Results}\label{sec:sec3}
In this section, we investigate the evolution of entanglement and nonlocality in two- and three-qubit systems for strong and ultrastrong coupling regimes, beginning with the initial W state. 
There are two primary reasons for choosing the W state as the initial three-qubit state.
Firstly, previous studies have demonstrated that the W and GHZ states have GTN and GTE both experimentally~\cite{PhysRevLett.129.060401,neeley2010generation} and theoretically~\cite{PhysRevA.81.052334,PhysRevLett.102.250404}.
However, most studies~\cite{PhysRevA.107.022201,wu2022genuine,wang2020genuine,wang2020violation} have only been able to analyze the GTN of general GHZ-like states using an exact expression for $\mathcal{N}(\rho_{abc})$.
Due to theoretical computational challenges, there are currently no specific analytical expressions available to calculate $\mathcal{N}(\rho_{abc})$ for general W states.
Thus, it is valuable to investigate how the GTN of W states evolves in real environments (dissipative systems) and whether there is a phenomenon of ``sudden death'' or ``sudden birth'' ~\cite{PhysRevA.109.012416} of GTN during the evolution in this context deserves further exploration.
Secondly, in the model we considered, when the coupling strength of all qubits to the bath is uniform (e.g., $\alpha_1 =\alpha_2=\alpha_3$), the probability amplitude $c_j (t)$ evolution remains consistent due to the symmetry of the W state.
However, what interesting physical phenomena can arise from inconsistent system-environment interactions due to varying coupling strengths?  Here, we provide clear answers to the questions.
\subsection{The cases of strong coupling regime}
\begin{figure}[htbp]
  \centering
  \includegraphics[width=4cm,height=3cm]{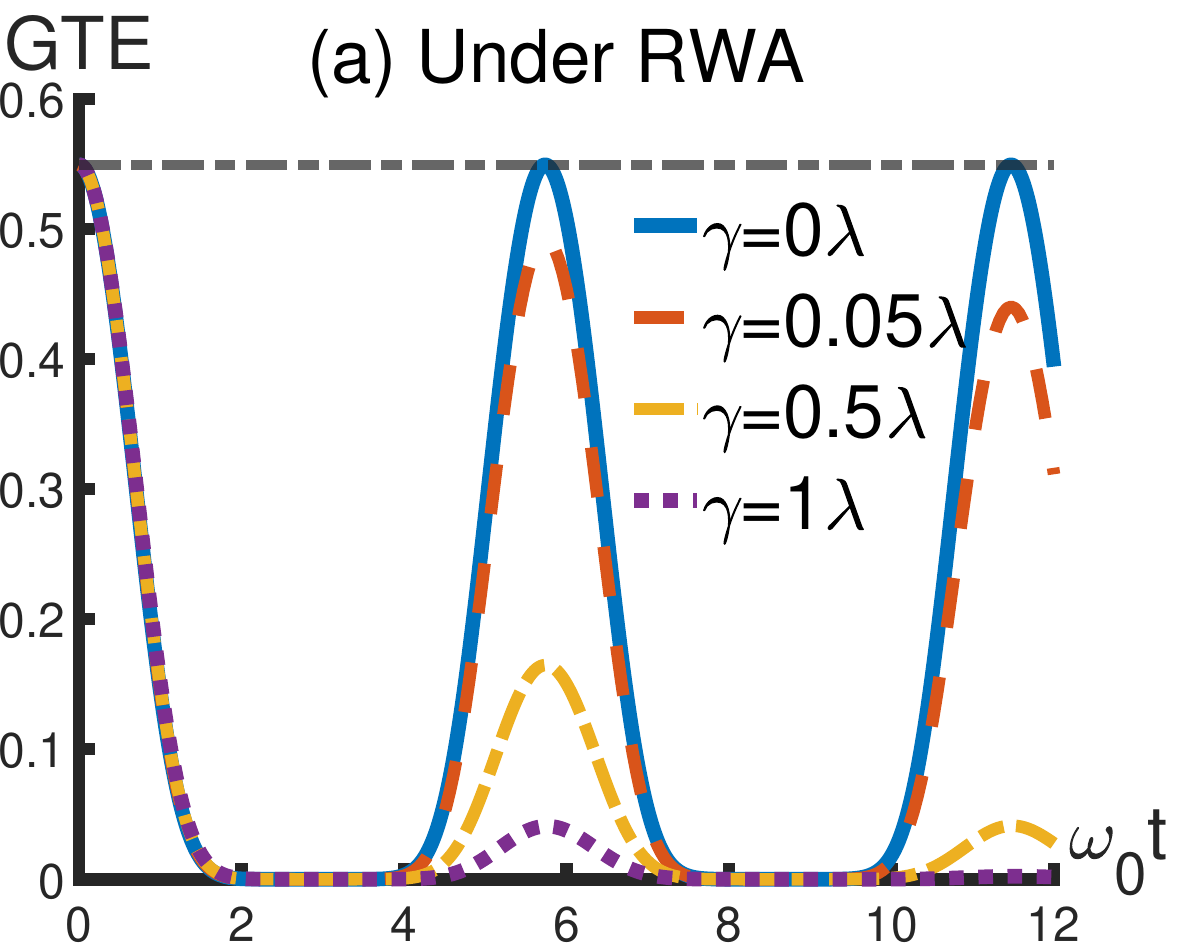}
  \includegraphics[width=4cm,height=3cm]{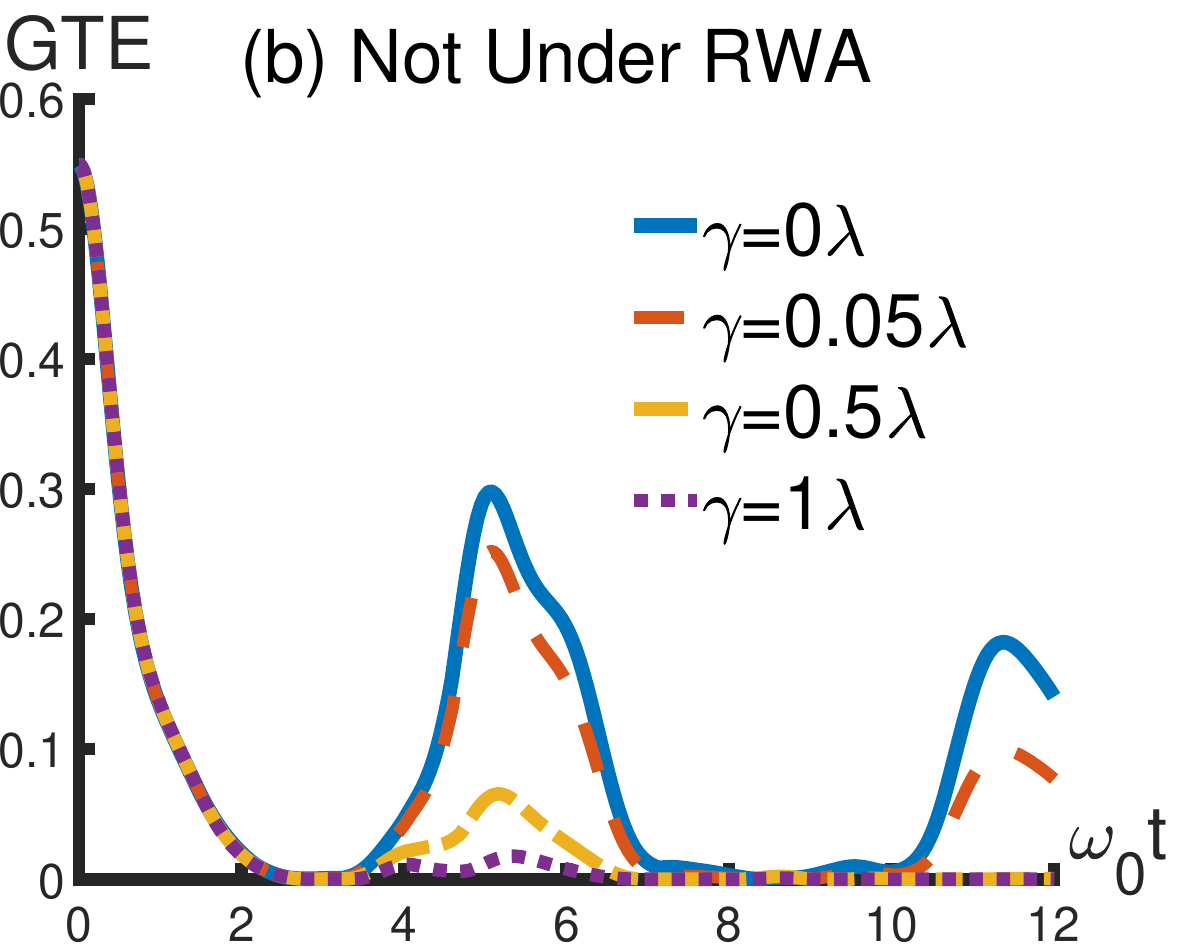}
  \includegraphics[width=4cm,height=3cm]{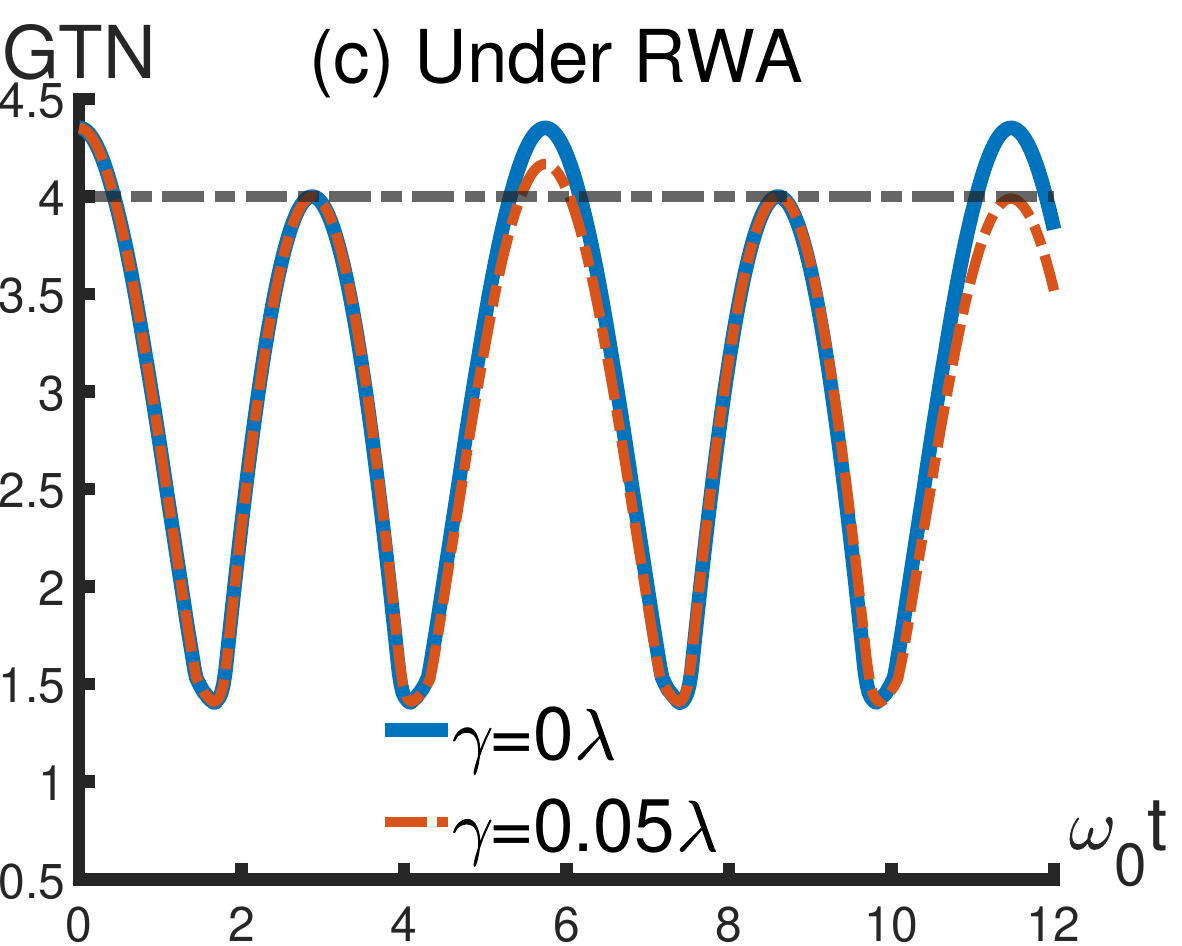}
  \includegraphics[width=4cm,height=3cm]{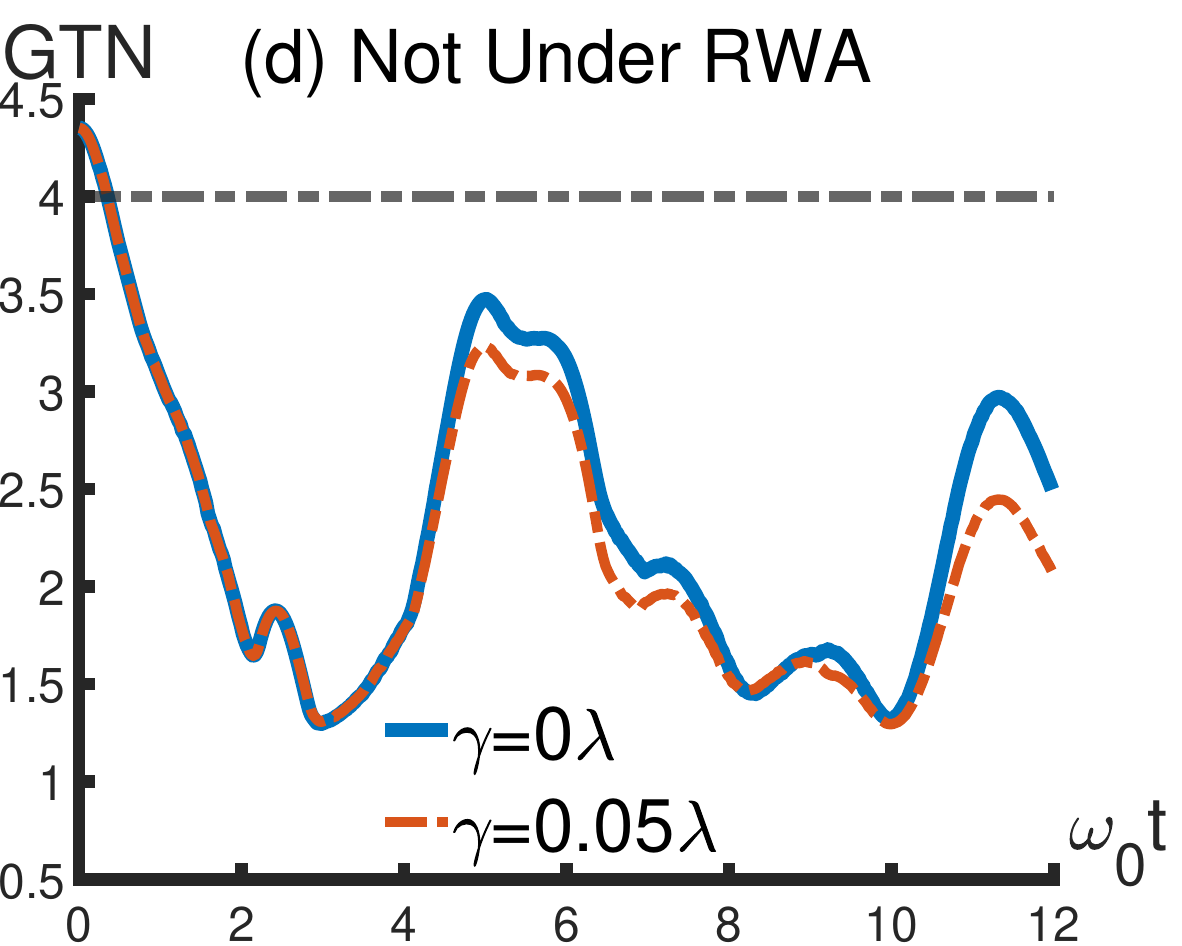}
  \caption{\label{fig:1}(a) GTE of the three-qubit system obtained by the exactly analytic expression with RWA for various values of $\gamma$; other parameters are chosen as $\alpha_1=\alpha_2=\alpha_3=1$, $\lambda=0.1\omega_0$, $\omega_0=1$ and the initial state is $\psi(0)=\ket{W}\otimes|0\rangle_{R}$.
  (b) The same as (a), but GTE obtained by the numerical HEOM method without RWA. (c) The same as (a), but GTN is quantified by calculating maximal violating of the Svetlichny inequality $\mathcal{N}(\rho_{abc})$, and no sudden birth of GTN is detected for the cases where $\gamma=0.5\lambda$ and $\gamma=1\lambda$. (d) The same as (c), but RWA is not applied here, and there is no sudden birth of GTN for all parameters $\gamma$.}
  \end{figure}
  In this subsection, we focus on the strong coupling regime ($\lambda=0.1\omega_0$) and compare the results obtained by the above methods for the initial W state.
  In Fig.~\ref{fig:1}, we present the GTE and GTN for the reduced three-qubit system. These results are obtained using the numerical HEOM method without RWA, as well as the exact analytic expression that applies RWA, detailed in Eq.~(\ref{Eq:RWA}).
  A comparison of Fig.~\ref{fig:1}(a) and Fig.~\ref{fig:1}(b) shows that the counter-rotating wave terms significantly accelerate the decoherence of GTE and suppress its sudden births. In Fig.~\ref{fig:1}(a), the GTE (blue solid line) displays a trend of periodic evolution that does not decay with $\gamma = 0$. This behavior occurs because the mode's lifetime $\tau_c = \frac{1}{\gamma}$ is infinite in the single-mode limit. In contrast, GTE clearly diminishes when RWA is not applied, as shown by the blue solid line in Fig.~\ref{fig:1}(b).
  As for GTN, we can find that the reduced three-qubit system violates the Svetlichny inequality with $\mathcal{N}(\rho_{abc})>4$ at the beginning, and the phenomenon of sudden death and sudden birth of GTN can be found due the memory effect of the bath in Fig.~\ref{fig:1}(c). However, in Fig.~\ref{fig:1}(d), the Svetlichny inequality is not violated in any cases except for the initial time interval.
  Consequently, the findings indicate that the counter-rotating wave terms diminish the revival amplitude of GTN and GTE, suggesting that the RWA fails to describe the real physical process.

 \begin{figure}[htbp]
  \centering
  \includegraphics[width=4cm,height=3cm]{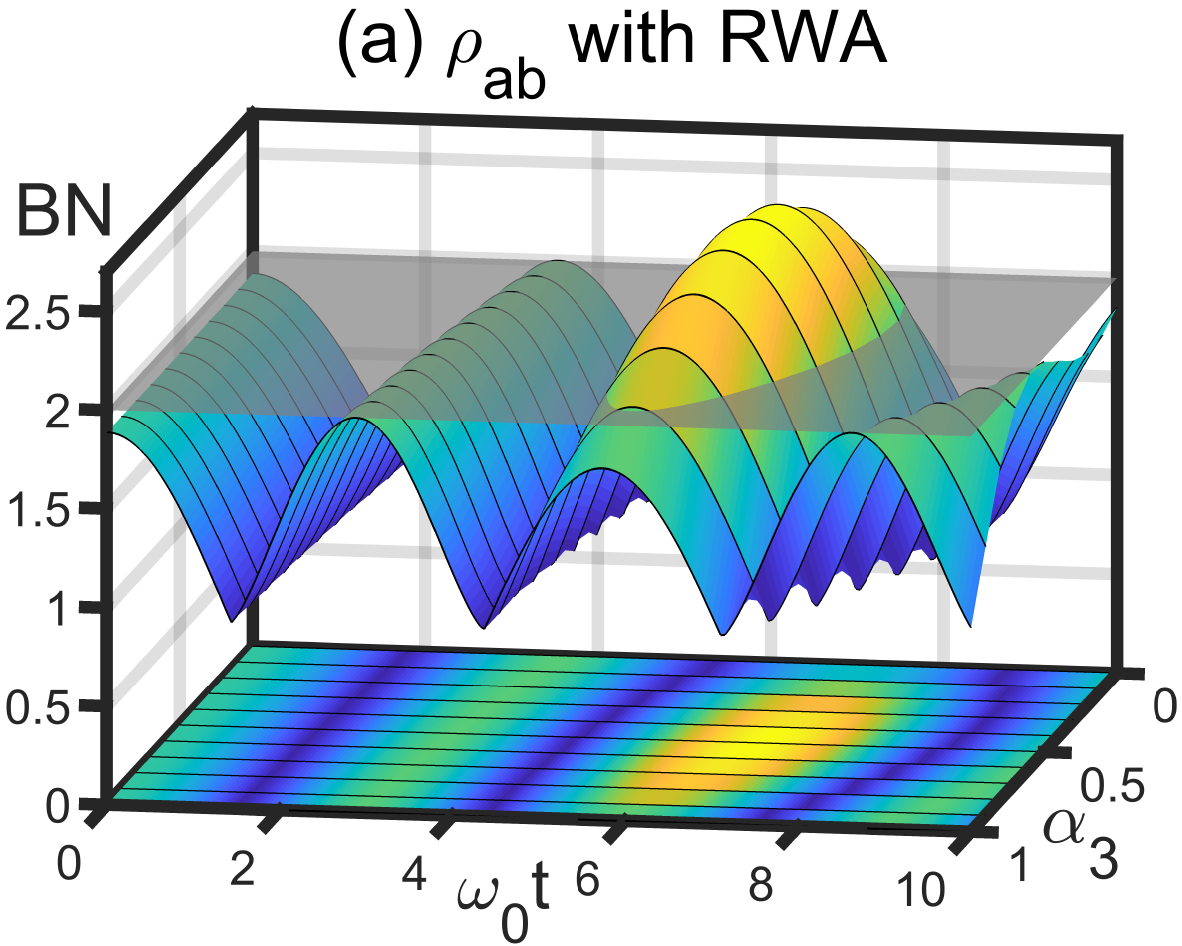}
  \includegraphics[width=4cm,height=3cm]{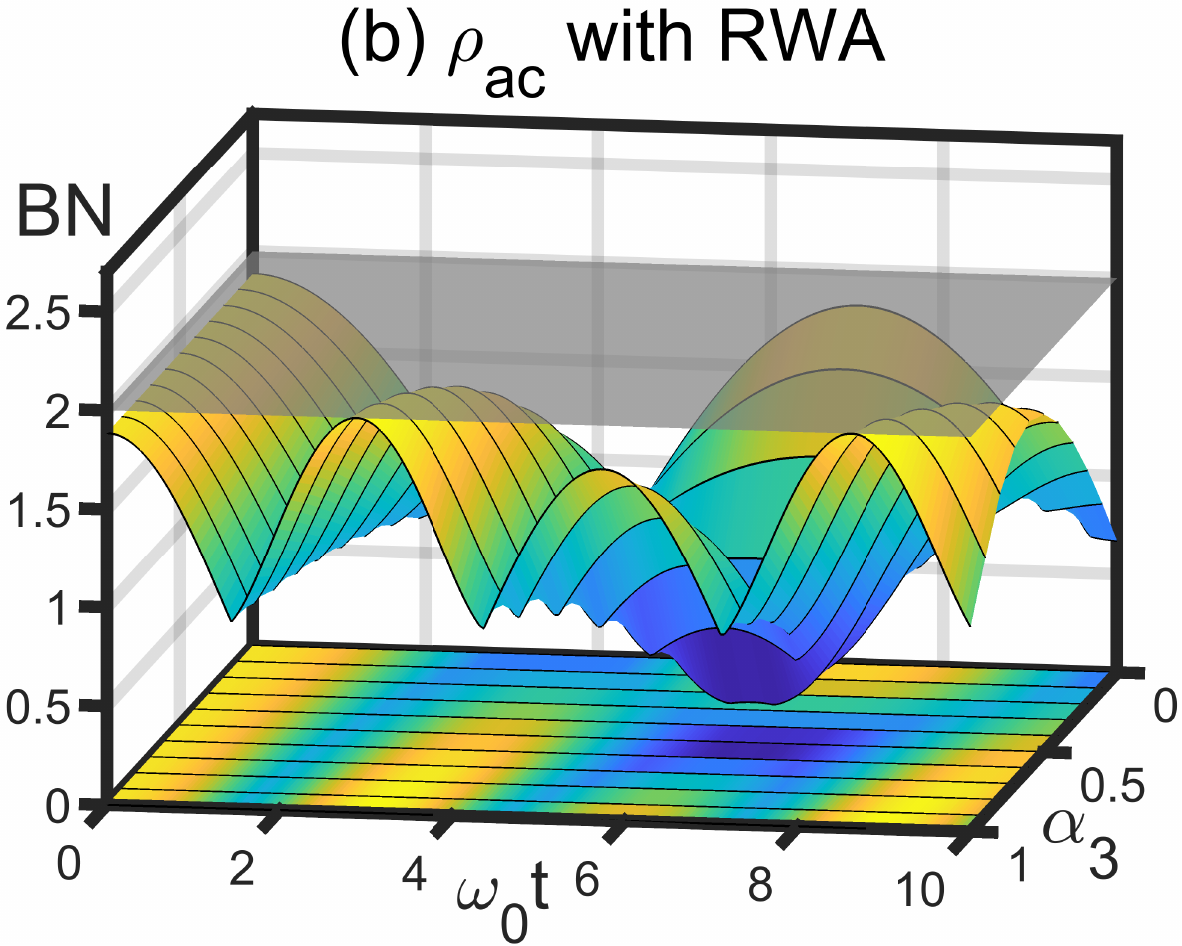}
  \includegraphics[width=4cm,height=3cm]{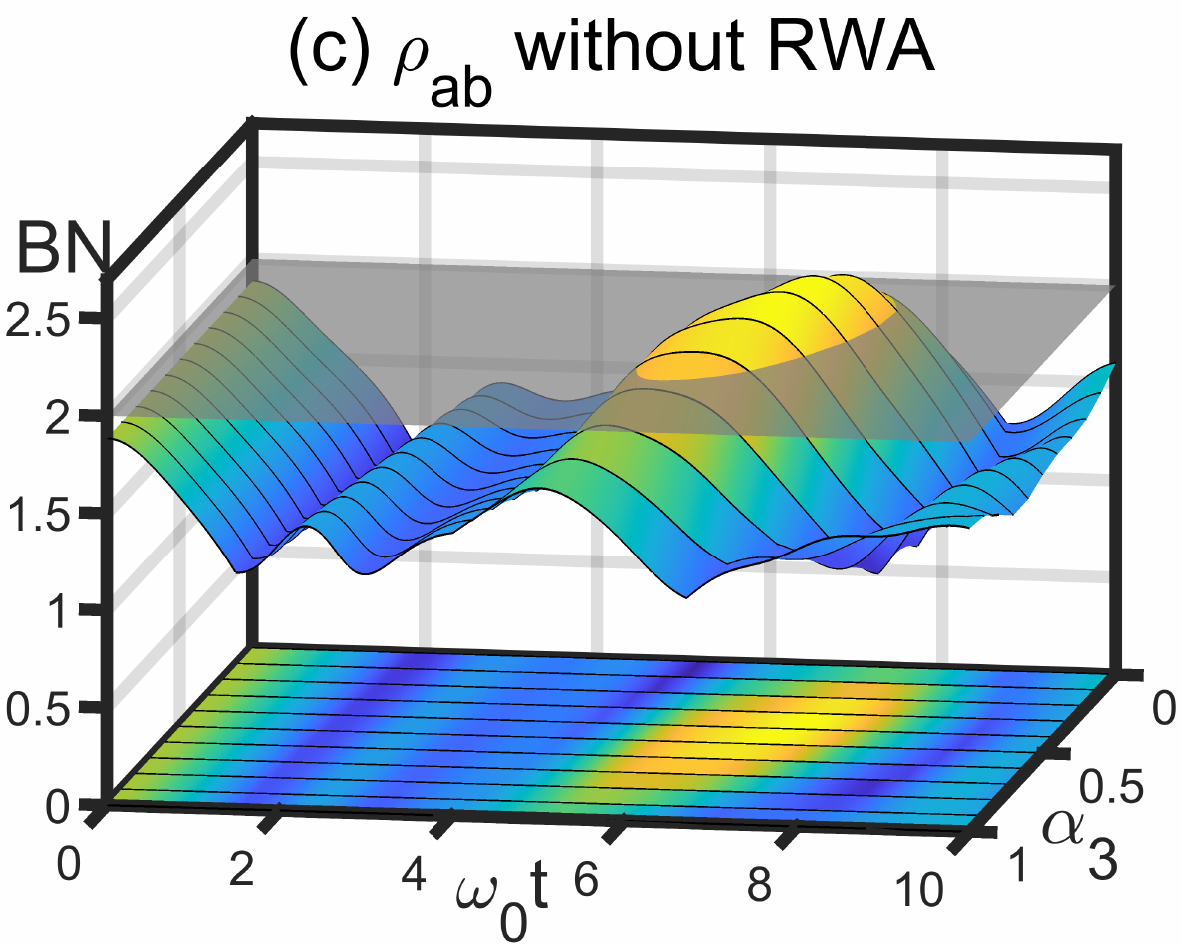}
  \includegraphics[width=4cm,height=3cm]{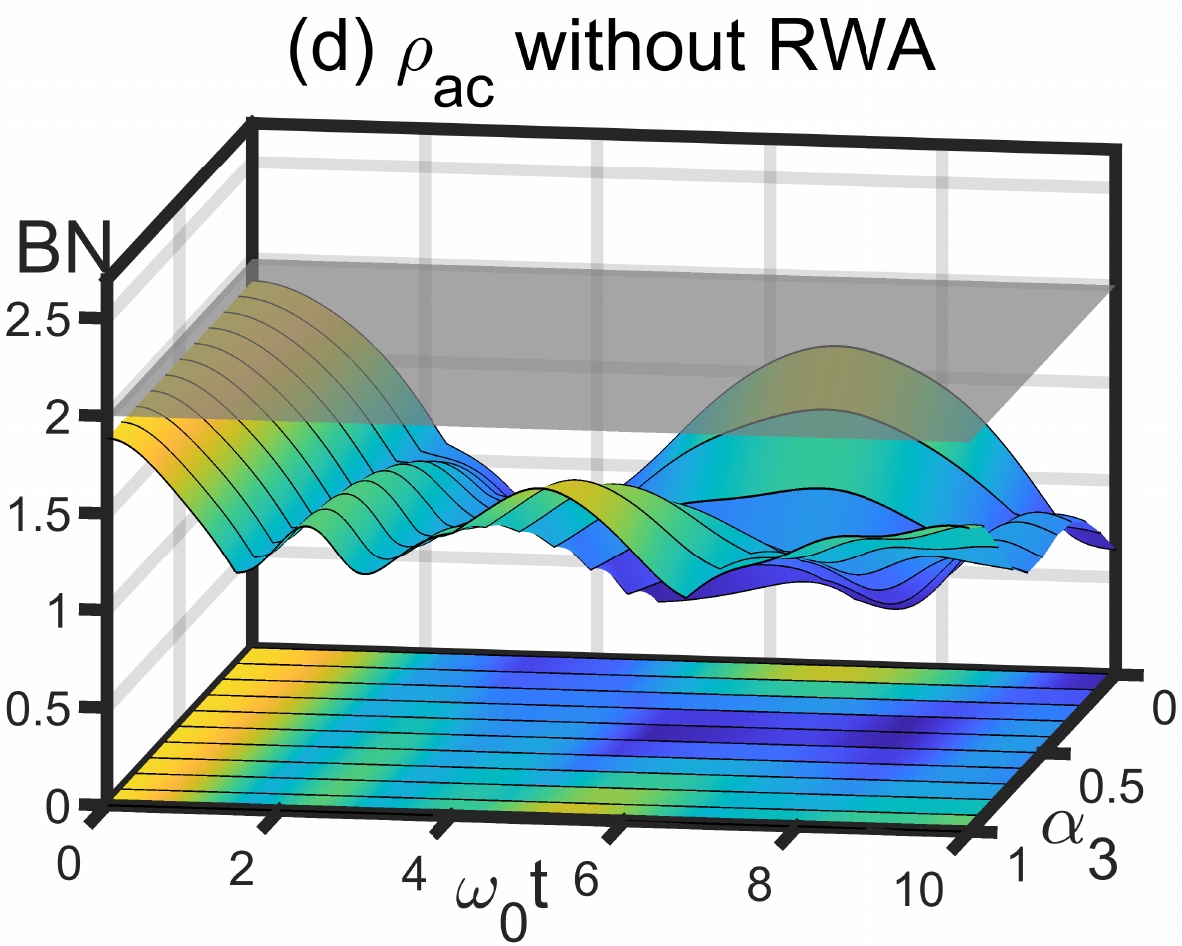}
  \caption{\label{fig:2} The BN is quantified by calculating maximal violating of the CHSH inequality for the subsystem $\mathcal{N}(\rho_{ab})$ and $\mathcal{N}(\rho_{ac})$, which is analyzed using both the exact analytic expression with RWA and the numerical HEOM method without RWA for different parameters: the coupling strength between the third qubit and the bath, denoted as \(\alpha_3\), varies from 0 to 1; other parameters are set as $\alpha_1=\alpha_2=1$, $\lambda=0.1\omega_0$, $\omega_0=1$ and the initial state is $\psi(0)=\ket{W}\otimes|0\rangle_{R}$. Due to the symmetry, we have $\rho_{ac}(t)=\rho_{bc}(t)$.
  In addition, the sudden birth of GTN is not observed in all cases.}
  \end{figure}
  Next, we examine the BN caused by inconsistent coupling strengths between the qubits and the bath. In this study, we adjust the coupling strength of one of the three qubits while keeping the other qubits unchanged. For simplicity, we assume that the coupling strengths between qubits and the bath are set as \(\alpha_1 = \alpha_2 \neq \alpha_3\).
  In Fig.~\ref{fig:2} (a) and (c), it is evident that the CHSH inequality is violated for the subsystem $\rho_{ab}$ when $\alpha_3 \neq \alpha_i$ (i=1,2), both with and without the RWA.
  The results also show that stronger BN is observed when the RWA is applied. 
   However, there is no violation of the CHSH inequality for the subsystem $\rho_{ac}$ and $\rho_{bc}$ throughout the time evolution.
   This observation is intriguing because it suggests that subsystems can exhibit nonlocal correlations due to the asymmetrical coupling between the qubits and the bath.
   Furthermore, although the three-qubit system initially exhibits GTN, there is no sudden birth of GTN during the subsequent evolution. Thus, nonlocality is transferred from the three-qubit system to its subsystems due to the interaction between the qubits and the bath.
   The results above raise two critical questions: (\romannumeral1).~Why does the asymmetric coupling between qubits and the bath result in BN? (\romannumeral2).~Why is the violation of the CHSH inequality more pronounced under the RWA?

\begin{figure}[htbp]
    \centering
    \includegraphics[width=4cm,height=3cm]{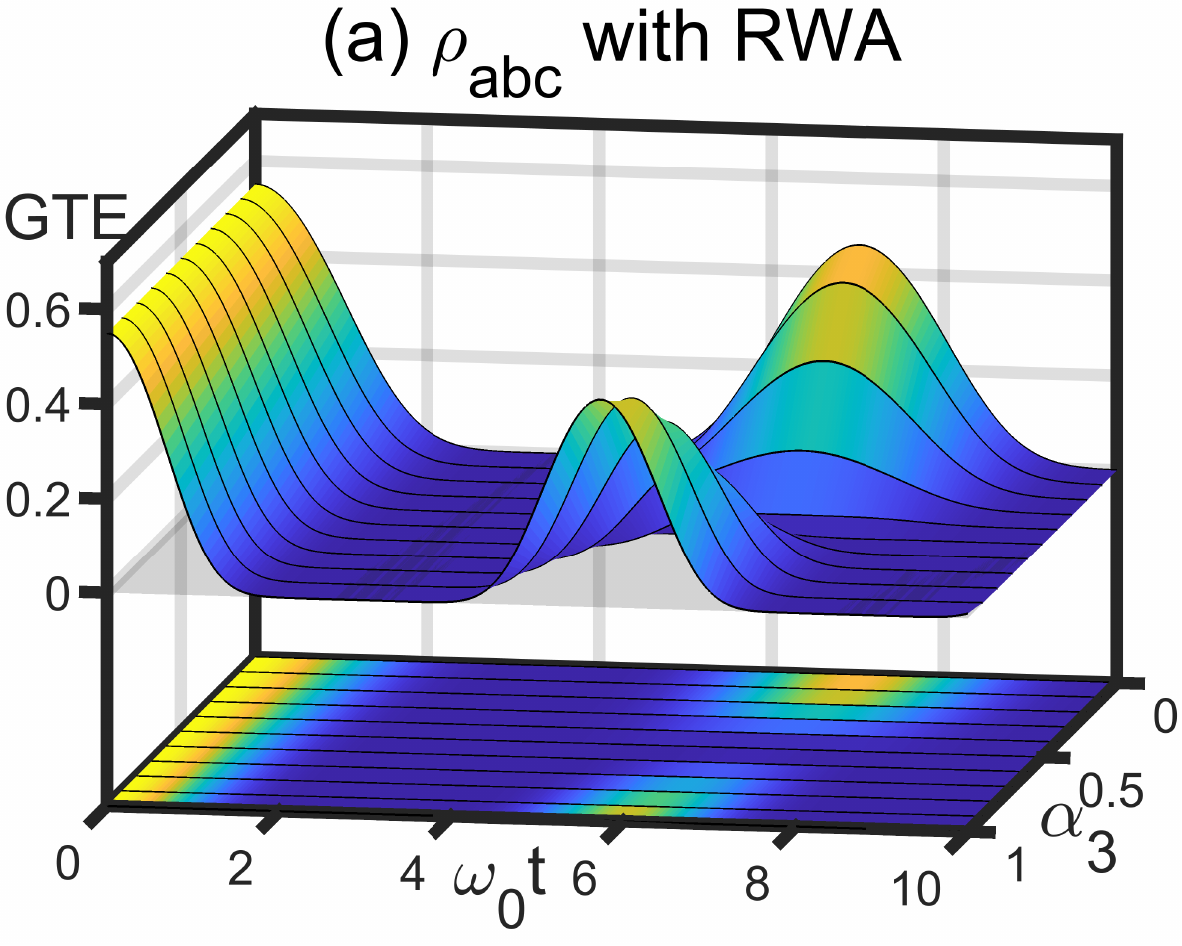}
    \includegraphics[width=4cm,height=3cm]{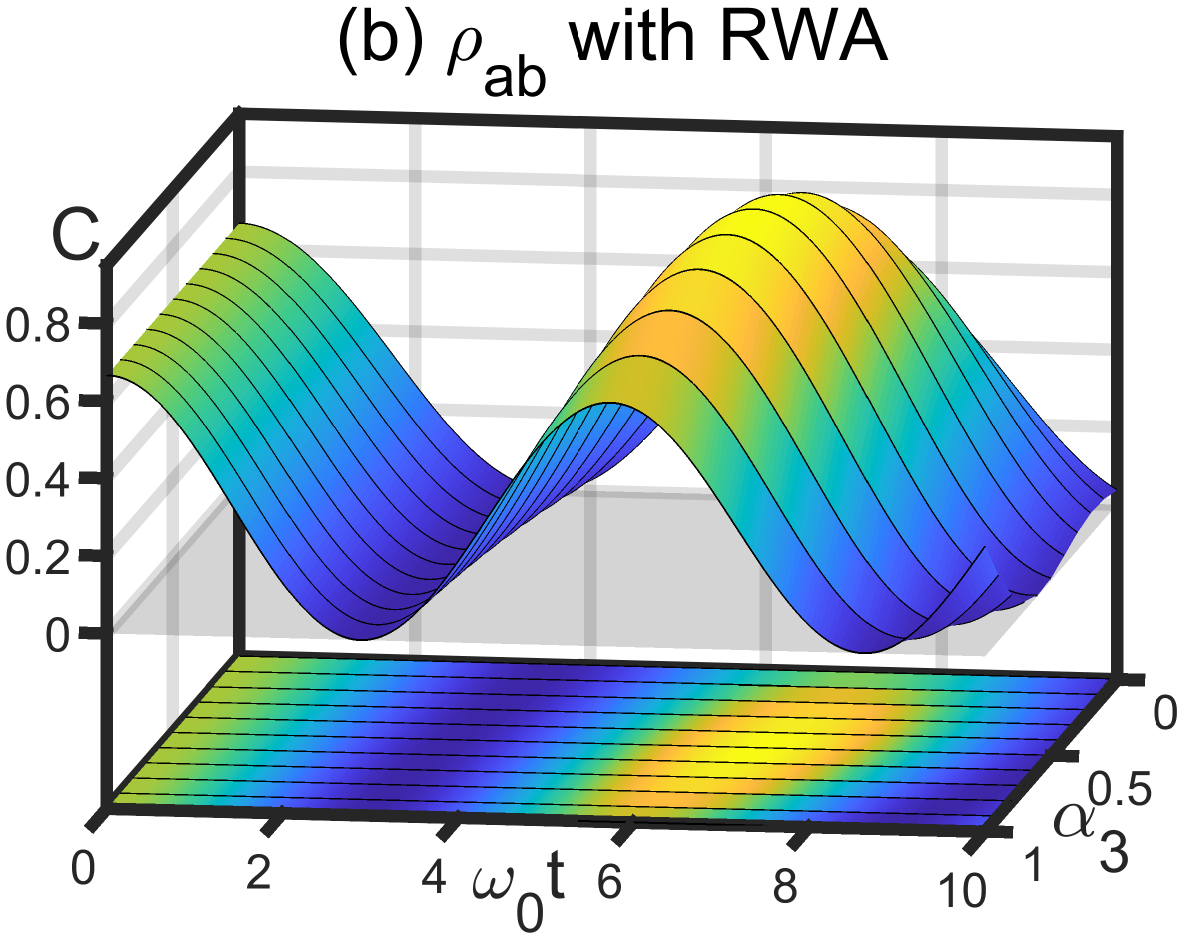}
    \includegraphics[width=4cm,height=3cm]{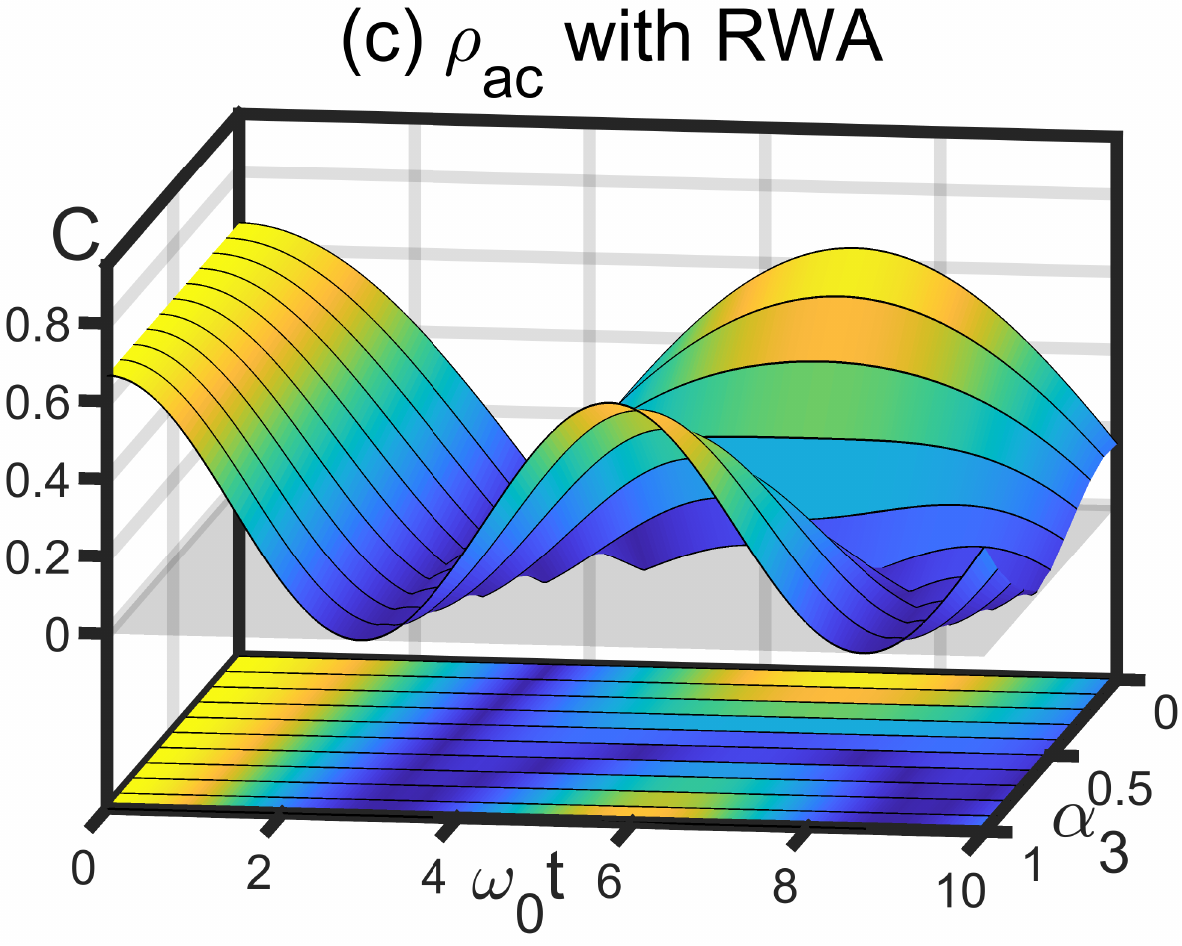}
    \includegraphics[width=4cm,height=3cm]{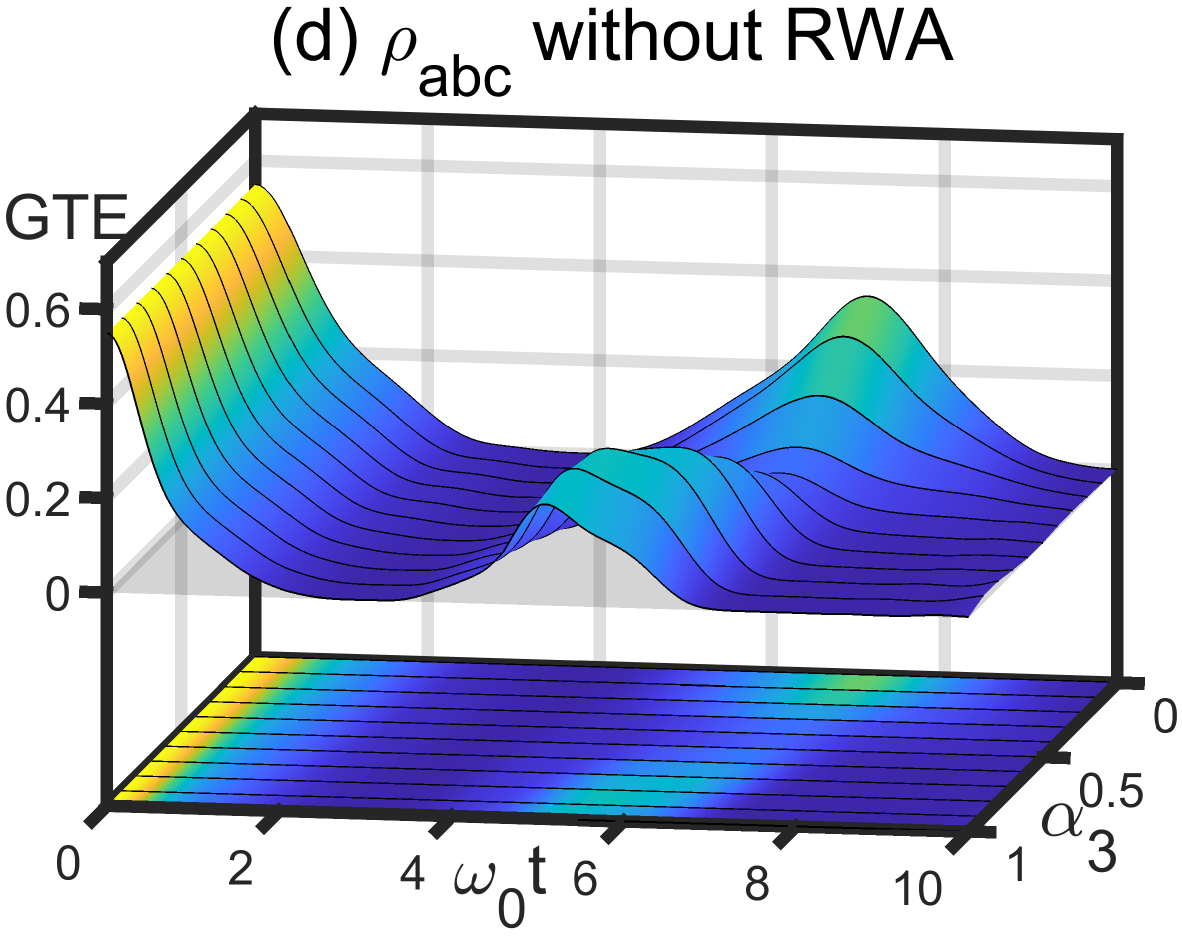}
    \includegraphics[width=4cm,height=3cm]{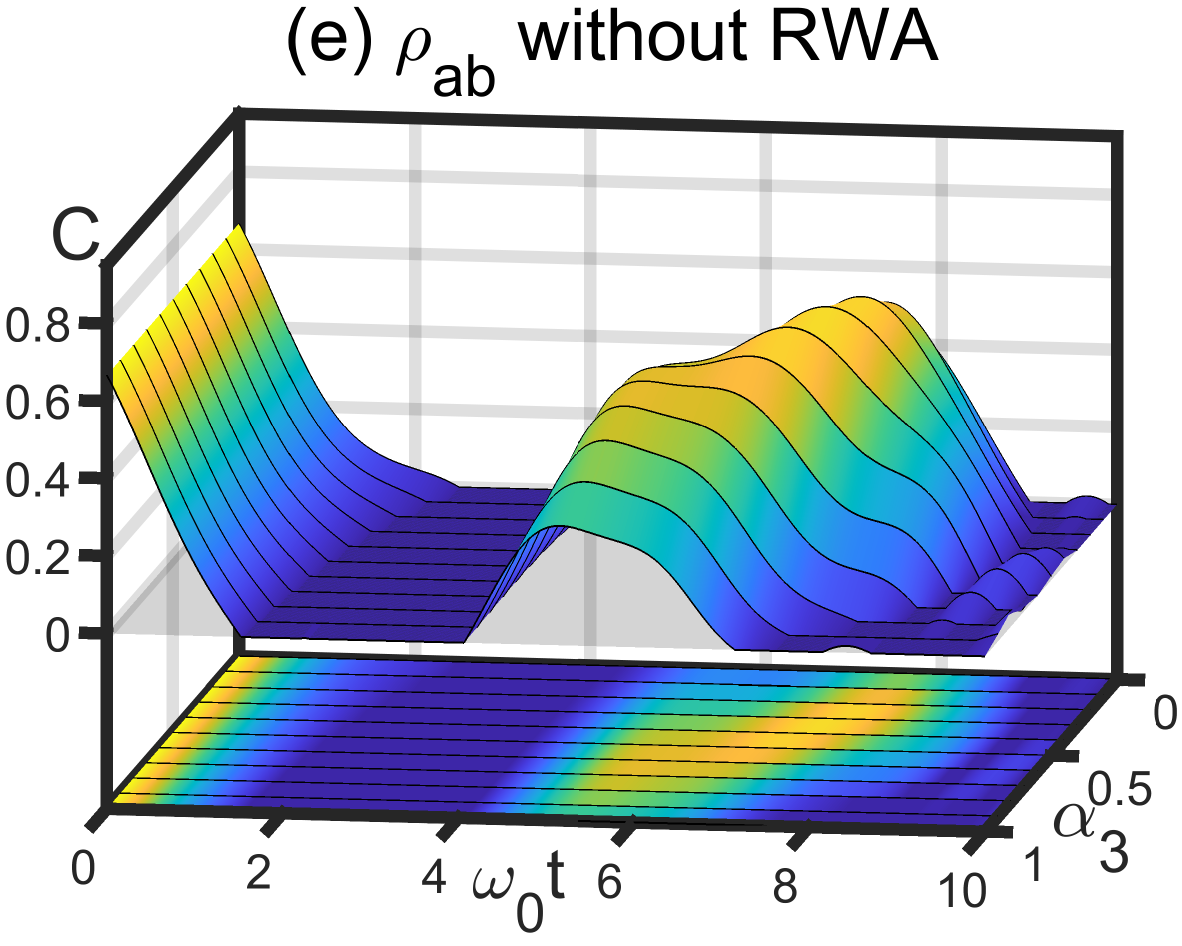}
    \includegraphics[width=4cm,height=3cm]{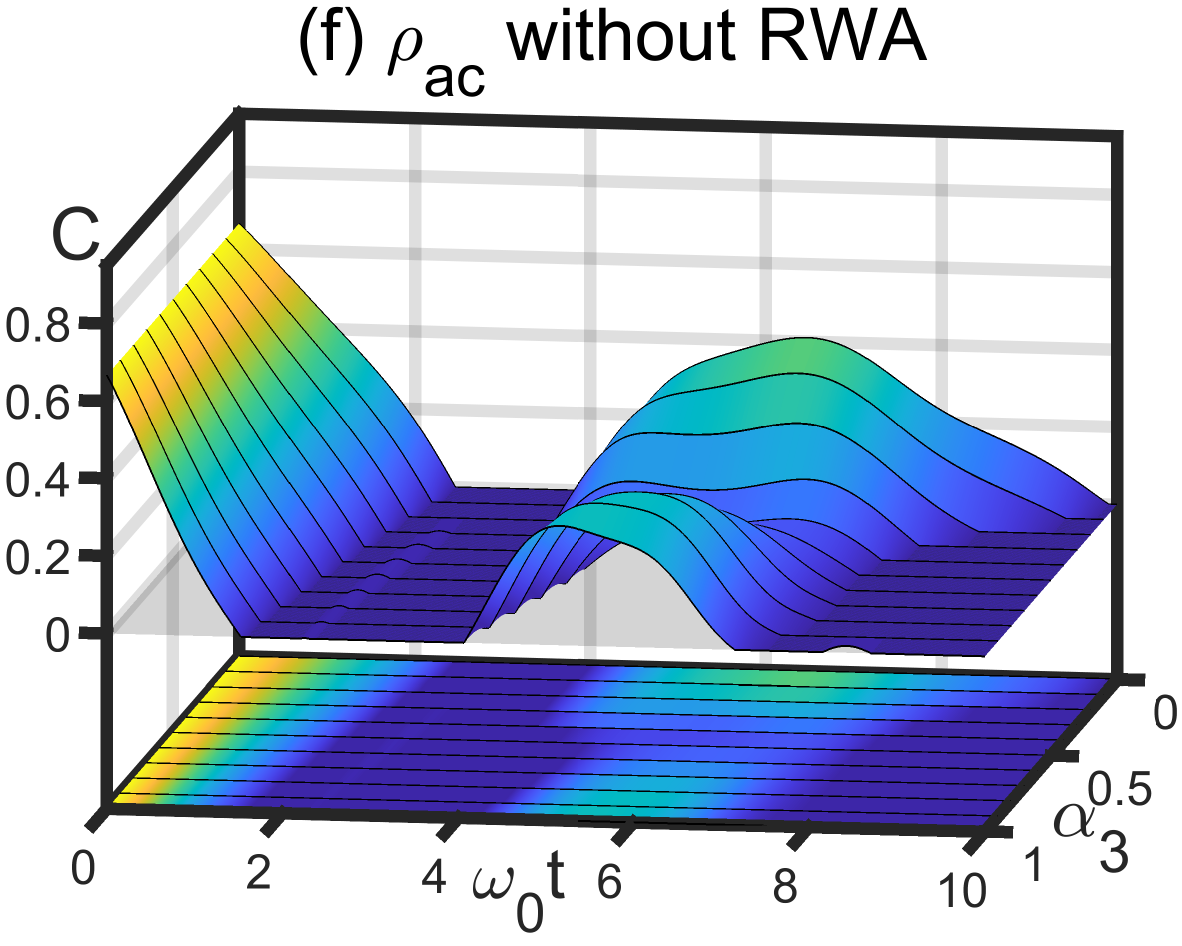}
    \caption{\label{fig:3} The dynamics of GTE and concurrence for the reduced three-qubit system $\rho_{abc}(t)$ and two-qubit subsystems $\rho_{ab}(t)$ and $\rho_{ac}(t)$, both with and without the RWA, are plotted as the function of $\omega_0 t$ for different parameters: $\alpha_3$ varies from 0 to 1; other parameters are set as $\alpha_1=\alpha_2=1$, $\lambda=0.1\omega_0$, $\omega_0=1$ and the initial state is $\psi(0)=\ket{W}\otimes  |0\rangle_{R}$.
    Besides, the sudden birth of GTN is not detected by using Svetlichny inequality throughout the time evolution of $\rho_{abc}(t)$ for all cases.}
    \end{figure}

   To explore these questions, we systematically investigate the entanglement in two- and three-qubit systems from the perspective of information flows. In Fig.~\ref{fig:3}, it can be observed that GTE and concurrence of the qubit systems decay rapidly to zero in both cases as the information from the qubit systems flows into the bath. However, the amount of information flowing back to the three qubits varies due to differences in coupling strength. This variation results in differing amplitude of concurrence revival for the states $\rho_{ab}$ and $\rho_{ac}$.
   More specifically, when comparing Fig.~\ref{fig:3} (b) and (c), it is evident that the revival of the concurrence for subsystem $\rho_{ab}$ caused by the bath's information flowing back (or memory effect) is significantly greater than that of subsystem $\rho_{ac}$ when the coupling strength $\alpha_3$ is around 0.5. This phenomenon occurs because when the information flows from the bath returns to the qubit system, more information is transferred back to subsystem $\rho_{ab}$ with $\alpha_1=\alpha_2 \textgreater \alpha_3$. 
  Next, we examine the impact of counter-rotating-wave terms on the time evolution of the states \(\rho_{abc}(t)\) and \(\rho_{ab}(t)\). As shown in Fig.~\ref{fig:3} (d)-(f), the revival amplitude of GTE and concurrence is significantly diminished under the non-RWA, which indicates that the counter-rotating-wave terms weaken the memory effect, leading to more information dissipated in the bath. 
  Our findings present a different result compared to previous research~\cite{PhysRevA.96.032125} and offer a fresh perspective on this topic.
\subsection{The cases of ultrastrong coupling regime}\label{sec:sec3b}
In this subsection, we study the impact of the counter-rotating wave terms on the qubit system when its coupling strength is ultrastrong with the bath.
\begin{figure}[htbp]
  \centering
  \includegraphics[width=4cm,height=3cm]{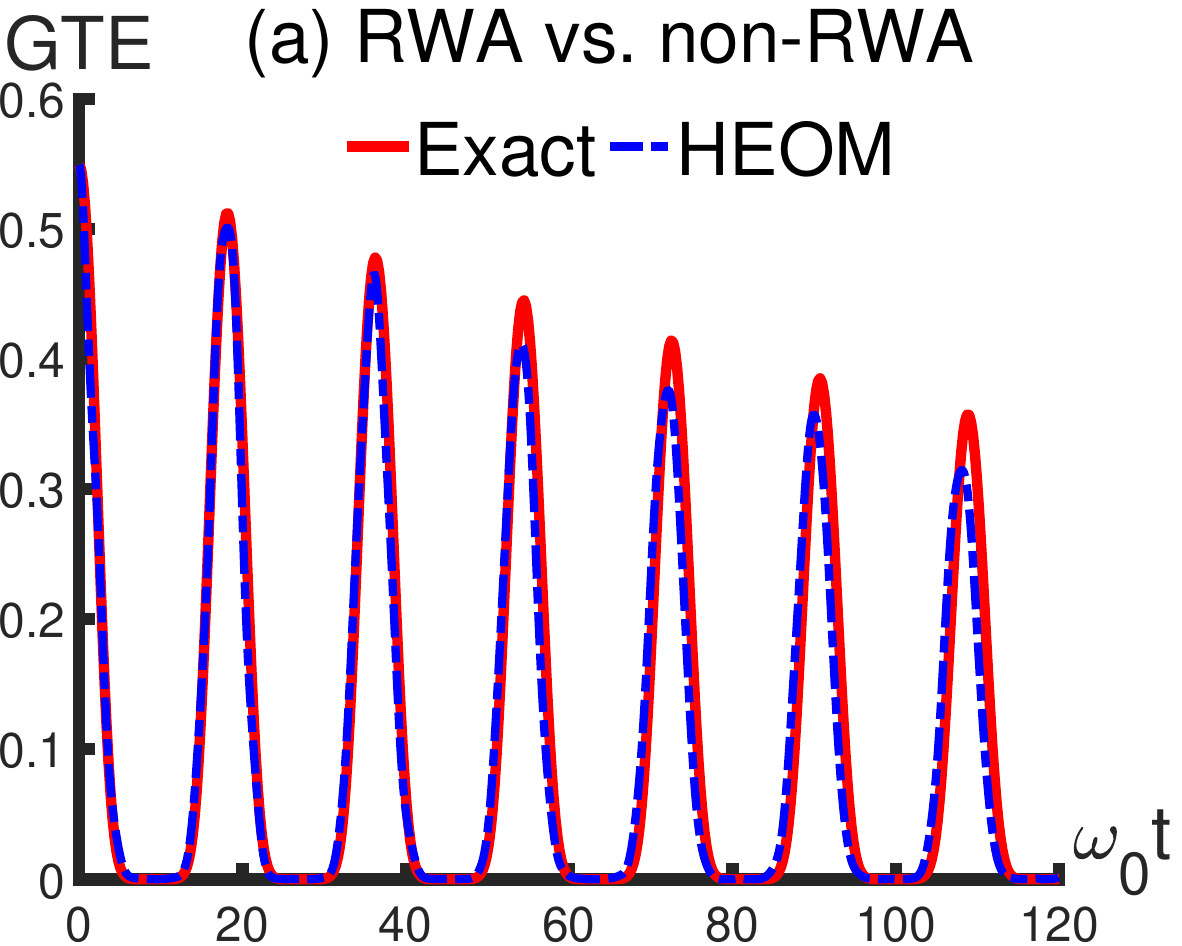}
  \includegraphics[width=4cm,height=3cm]{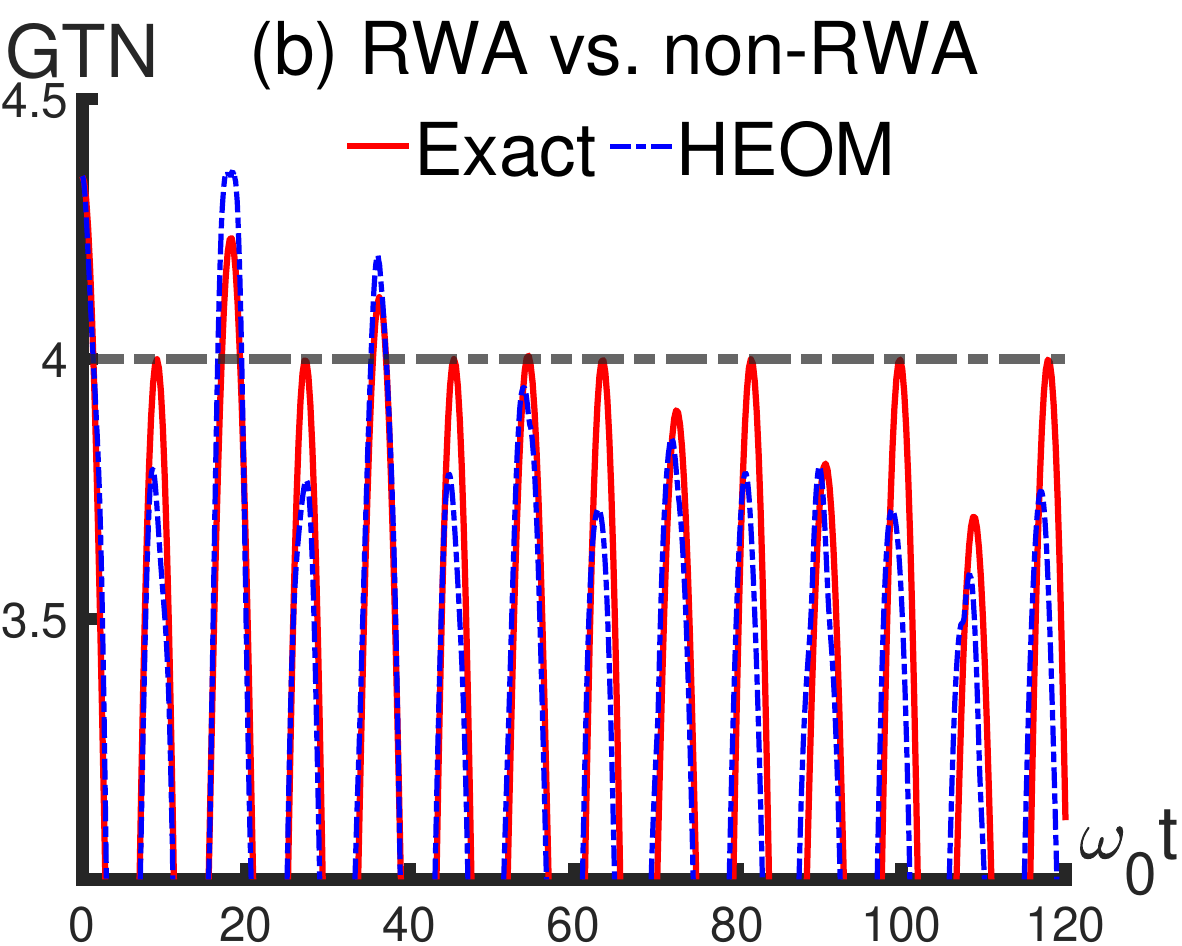}
  \caption{\label{fig:3b1}(a) GTE of the three-qubit system is obtained using the numerical HEOM method without the RWA and the exact analytic expression with the RWA for ultrastrong coupling regime \(\lambda = 0.01\omega_0\). The parameters are set as follows: \(\gamma = 0.01\lambda\), \(\omega_0 = 1\), and \(\alpha_1 = \alpha_2 = \alpha_3 = 1\). The initial state is defined as $|\psi(0)\rangle=\ket{W}\otimes  |0\rangle_{R}$. 
  (b) the same as (a), but here the GTN is calculated by calculating maximal violating of the Svetlichny inequality $\mathcal{N}(\rho_{abc})$. Besides, there is no sudden birth of BN for any subsystem.}
  \end{figure}
  In Fig.~\ref{fig:3b1}(a), one can find that GTE can be effectively revived due to the memory effect in both RWA and non-RWA cases, and the counter-rotating-wave terms only limitedly weaken the revival amplitude of GTE, which shows that the dynamic evolution is quite different compared with strong coupling cases.
  In Fig.~\ref{fig:3b1}(b), the GTN also shows a series of sudden deaths and births phenomenon. However, the counter-rotating wave terms significantly enhance the sudden birth amplitude of GTN, which suggests that the virtual excitations caused by these counter-rotating wave terms have contrasting influences on different quantum correlations.

  \begin{figure}[htbp]
    \centering
    \includegraphics[width=6cm,height=5cm]{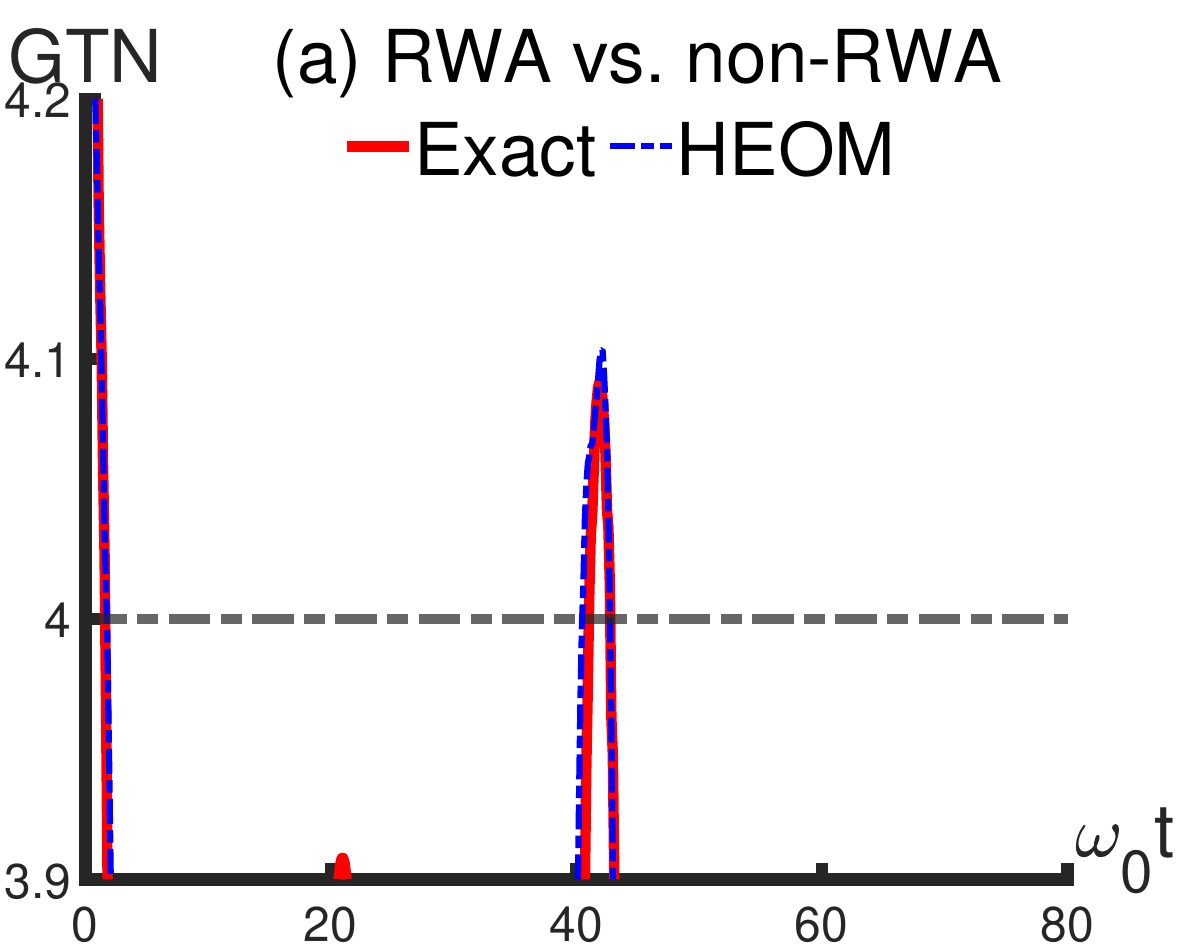}
    \includegraphics[width=6cm,height=5cm]{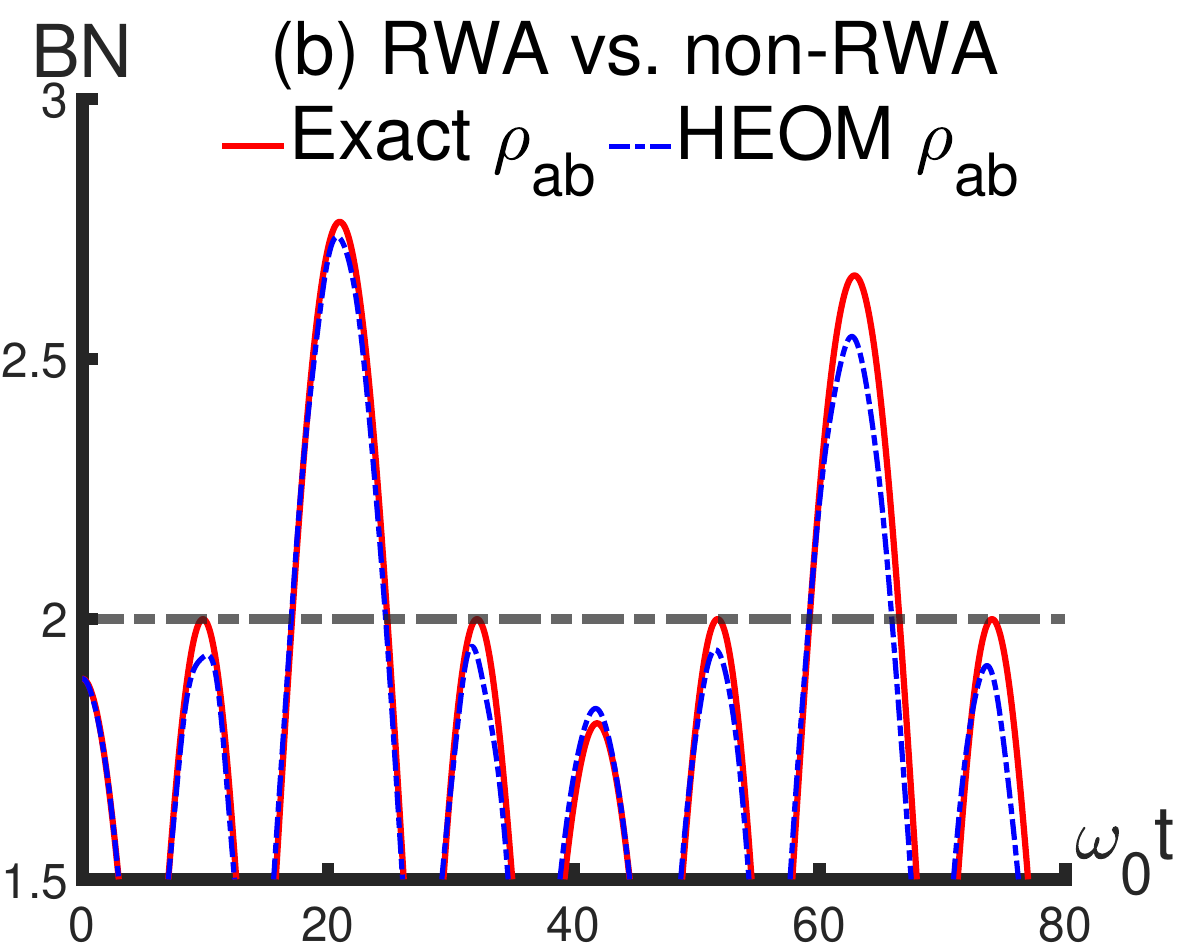}
    \caption{\label{fig:3b2}(a) GTN of the three-qubit system is obtained by calculating maximal violating of the Svetlichny inequality $\mathcal{N}(\rho_{abc})$ with and without the RWA.
    The parameters are set as follows: \(\gamma = 0.01\lambda\), \(\omega_0 = 1\), $\alpha_1 = \alpha_2=1$  and \(\alpha_3 = 0.5\). 
    The initial state is defined as $|\psi(0)\rangle=\ket{W}\otimes  |0\rangle_{R}$. 
    (b) the same as (a), but here the maximal violating of the CHSH inequality $\mathcal{N}(\rho_{ab})$ is calculated for subsystem $\rho_{ab}$. Due to symmetry, we have $\rho_{ac}(t)=\rho_{bc}(t)$. The BN is not observed for subsystems $\rho_{ac}(t)$ and $\rho_{bc}(t)$ throughout the time evolution.}
  \end{figure}
  In the preceding subsection, we discussed the phenomenon of nonlocality transfer from the three-qubit system \(\rho_{abc}\) to its subsystem \(\rho_{ab}\) in Fig.~\ref{fig:2}. It is important to note that the sudden birth of GTN is not observed in these cases. However, when we focus on the ultrastrong coupling regime, we can see the transfer of nonlocality back and forth in both the three-qubit and two-qubit systems.
  In Fig.~\ref{fig:3b2} (a) and (b), both the Svetlichny inequality and the CHSH inequality are violated at different time intervals. Specifically, the three-qubit system demonstrates GTN while the subsystem is local at \(\omega_0 t = 0 \). The CHSH inequality is violated around \( \omega_0 t = 20 \), and GTN disappears simultaneously. This suggests that nonlocality has transitioned from the three-qubit system to the subsystem \( \rho_{ab} \) due to the interaction between the qubit system and bath.
Moreover, the Svetlichny inequality is violated around \(\omega_0 t = 40 \), without detectable two-party nonlocality, which indicates a transfer of nonlocality from the subsystem back to the overall system. Finally, the GTN vanishes, and nonlocality in the two-qubit subsystem \( \rho_{ab} \) revival around \(\omega_0 t = 60 \).
It is crucial to note that our findings regarding the nonlocality of the two- and three-qubit systems conform to the monogamy and complementarity relationships of multi-party nonlocality as discussed in \cite{PhysRevA.96.022121}. These relationships were introduced by Anubhav Chaturvedi et al. and analyzed through the nonlocal game. Here, we present a numerical case to verify these relationships from the perspective of dynamic evolution.
   
\begin{figure}[htbp!]
    \centering
    \includegraphics[width=4cm,height=3cm]{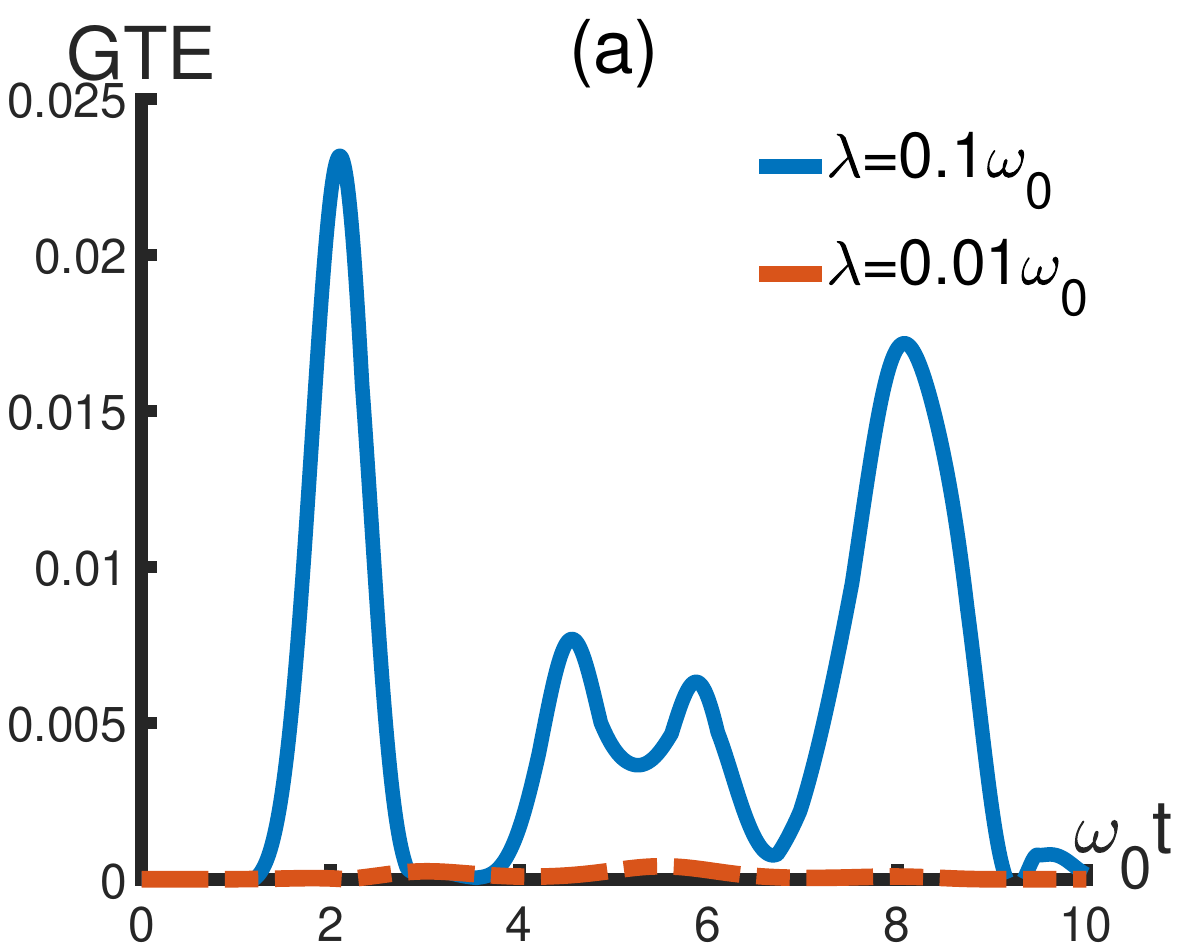}
    \includegraphics[width=4cm,height=3cm]{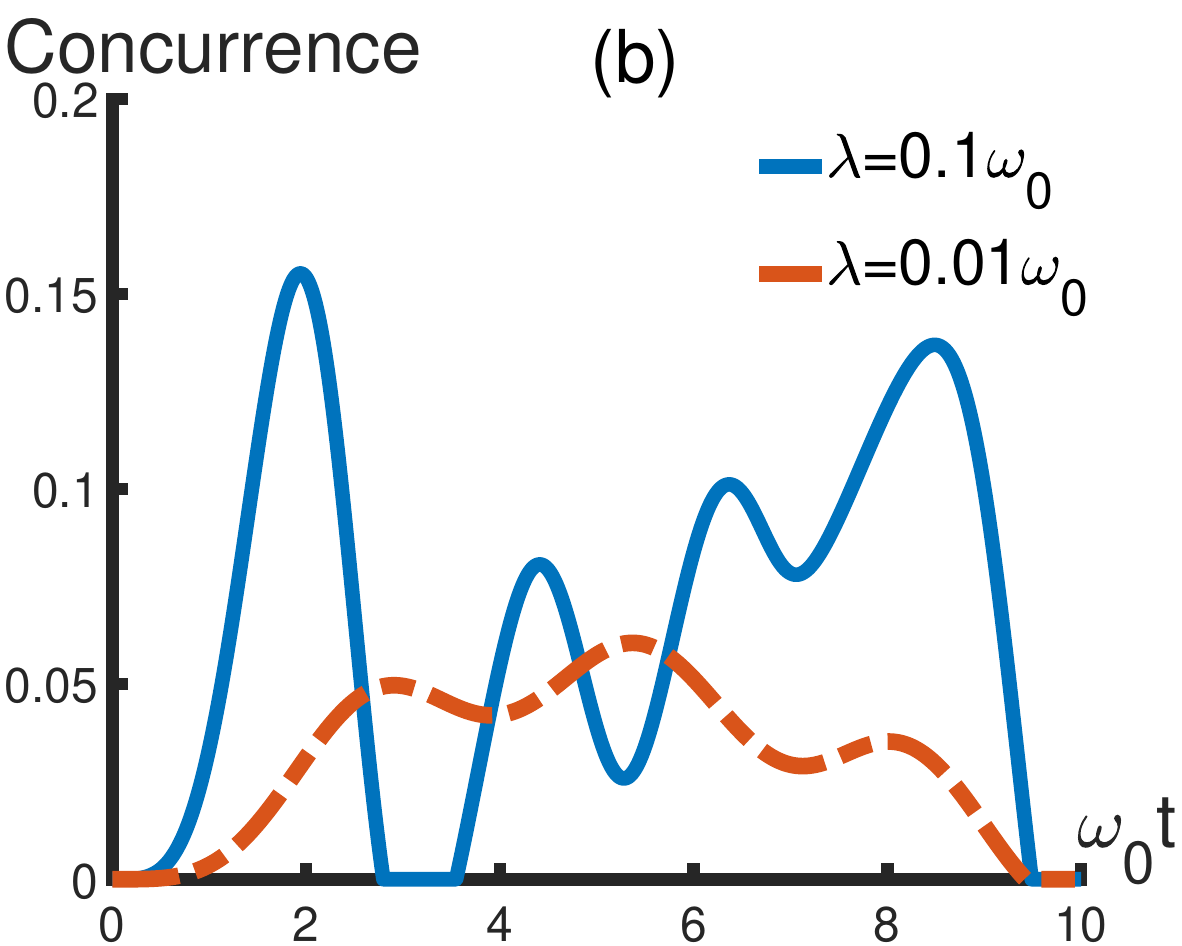}
    \caption{\label{fig:3b.3}(a) GTE of the three-qubit system is obtained using the numerical HEOM method without RWA both for strong ($\lambda=0.1\omega_0$) and ultrastrong coupling ($\lambda=0.01\omega_0$) regimes.
    The initial state is defined as $|\psi(0)\rangle=\ket{ggg}\otimes  |0\rangle_{R}$ and the parameters are set as follows: $\gamma = 0.01\omega_0$, $\omega_0 = 1$, and $\alpha_1 = \alpha_2 = \alpha_3 = 1$ . 
        (b) the same as (a), but here the concurrence is obtained for subsystem $\rho_{ab}$, due to the symmetry, we have $\rho_{ab}(t)=\rho_{ac}(t)=\rho_{bc}(t)$.
        Under the RWA, the initial state does not evolve; both GTE and concurrence stay at zero. Additionally, there is no sudden birth of GTN and BN in three- and two-qubit systems in any case.}
    \end{figure}

  In previous studies, researchers have focused on generating quantum correlations from the initial state \( |\psi(0)\rangle = \ket{gg} \otimes |0\rangle_R \). 
  Although this type of state does not evolve under the RWA, results indicate that quantum entanglement and steering can be generated through the virtual excitations caused by the counter-rotating 
  wave terms~\cite{PhysRevA.85.062323,PhysRevA.97.052309,ALTINTAS20121791}.
  Naturally, here we have a question: Does virtual excitation have the same generative effect on genuine three-party correlations?
  Therefore, we investigate the effects of counter-rotating wave terms on the GTE and the concurrence of the initial state here.
  In Fig.~\ref{fig:3b.3}, for strong coupling cases, one can find that the counter-rotating wave terms can induce GTE and concurrence of the subsystem, but the generated GTE is very weak.
   For ultrastrong coupling regime, while the concurrence of the subsystem can be generated, GTE is nearly zero during the time evolution. The results suggest that the counter-rotating-wave terms cannot induce entanglement in a three-party system as strongly as it does for a two-party system. Therefore, the conclusion of quantum correlation between two-party systems cannot be generalized to three-party scenarios.
\section{Conclusion}\label{sec:sec4}
In summary, we investigate the impact of counter-rotating-wave terms on the entanglement and nonlocality for initial $\ket{W}$ and $\ket{ggg}$ qubit state in quantum open systems by comparing the result between the non-RWA and RWA. In non-RWA cases, we use the HEOM approach to study the dynamics of the reduced three-qubit systems. This numerical technique is applicable when the system and the bath are coupled strongly without the treatment of Born-Markovian, perturbative, and rotating-wave approximations. We also obtain the exact time evolution equation of the reduced three-qubit system when RWA is applied.
In a comparison of the dynamic evolution of the two methods, the findings are as follows: 
(\romannumeral1). For the strong coupling regime, the counter-rotating wave terms can suppress the sudden birth amplitude of entanglement and nonlocality in both two- and three-qubit systems. They also accelerate the decoherence process of qubit systems, but the RWA fails to predict the real physical process.
(\romannumeral2). For the strong coupling regime, the sudden birth of BN is observed during the evolution of the two-qubit state when the coupling strength between each qubit and the bath is unequal. The physical mechanism of those phenomena is that the dissipated information can return to the qubit more efficiently with higher coupling strength.
(\romannumeral3). For the ultrastrong coupling regime, the counter-rotating wave terms significantly enhance the sudden birth amplitude of GTN but weaken that of GTE, which indicates that the counter-rotating wave terms play distinctly different roles in various quantum correlations.
(\romannumeral4). The new phenomenon of nonlocality transferring between two- and three-qubit systems is observed, fulfilling the complementarity and monogamy of multi-party nonlocality from the perspective of physical dynamics in the ultrastrong coupling regime.
(\romannumeral5). For the zero-excitation cases, the counter-rotating wave terms can generate concurrence in the two-qubit system, which is similar to previous studies~\cite{PhysRevA.85.062323,PhysRevA.97.052309,ALTINTAS20121791}. However, the induced GTE is quite weak in both strong and ultrastrong coupling regimes, suggesting that virtual excitations are limited in generating genuine three-party correlations.
Finally, the recent advancements in experiments related to preparing the various quantum states~\cite{kang2016fast,wei2015preparation,neeley2010generation} and engineering strong coupling in superconducting circuits and atom-cavity coupled systems~\cite{you2011atomic,niemczyk2010circuit,frisk2019ultrastrong} have led us to believe that our findings will be valuable for various experimental applications in quantum computing and quantum information processing.

\begin{acknowledgments}
  This work was supported by the National Natural Science Foundation of China under Grant Nos.11364006 and 11805065.
  \end{acknowledgments}

\appendix
\section{HEOM}\label{sec:secapp1}
In this appendix, we briefly outline the steps for using HEOM to analyze the reduced dynamics of an arbitrary three-qubit system interacting with a bosonic bath. For a generalized quantum dissipative system resulting from interactions with heat baths composed of harmonic oscillators, the total Hamiltonian can be expressed as follows~\cite{10.1063/5.0011599}:
\begin{align}\label{Eq:A1}
 \hat{H}_{\text {tot}} & =\hat{  H}_{S}+\hat{ H}_{I+B} \notag \\
 \hat{H}_{I+B}&=\sum_{a} \sum_{j=1}^{N_{a}}\left[\frac{\left(\hat{p}_{j}^{a}\right)^{2}}{2 m_{j}^{a}}+\frac{1}{2} m_{j}^{a}\left(\omega_{j}^{a}\right)^{2}\left(\hat{x}_{j}^{a}\right)^{2}-\alpha_{j}^{a} \hat{V}^{a} \hat{x}_{j}^{a}\right]
  \end{align}
  where \( \hat{H}_{S} \) refers to the Hamiltonian of the system. The term \( \hat{H}_{I+B} \) includes both the bath and the system-bath interaction Hamiltonian. The degrees of freedom of the bath, denoted as \(\alpha\), are represented as \( N_{\alpha} \) harmonic oscillators. Specifically, the \(\alpha\)th heat bath is characterized by the following parameters: momentum \(\hat{p}_{j}^{a}\), position \(\hat{x}_{j}^{a}\), frequencies \(\omega_{j}^{a}\), coupling coefficients \(\alpha_{j}^{a}\), and masses \(m_{j}^{a}\). Additionally, both \( \hat{H}_{S} \) and \( \hat{V}^{a} \) can be expressed in the following form within the discretized energy spaces
\begin{align}
&  \hat{V}^{a}=\sum_{j, k} \hat{V}_{j k}^{a}|j\rangle\langle k|\\
&\hat{H}_{A}  = \sum_{j} \hbar \omega_{j}|j\rangle\langle j|+\sum_{j \neq k} \hbar \Delta_{j k}(t)|j\rangle\langle k|
\end{align}
here $\ket{j}$ is defined as the $j^{th}$ eigenenergy states of ${H}_{S}$. The spin-boson model we studied is a simplified version with the Eq.~(\ref{Eq:A1}), with $\alpha=1$ and $j=0,1$.

Then, the exact solution of the reduced quantum subsystem, including all orders of the system-bath interactions, can be derived by making use of the superoperator technique, which can be derived as~\cite{doi:10.1143/JPSJ.75.082001}
\begin{align}\label{Eq:appendix1}
  \hat{\rho}_{S}^{I}(t)  =& \hat{\mathcal{T}} \exp \big\{ -\int_{0}^{t} d t_{2} \int_{0}^{t_{2}} d t_{1} 
  \hat{ V}\left(t_{2}\right)^{\times} \big[ C^{R}\left(t_{2}-t_{1}\right) \hat{V}\left(t_{1}\right)^{\times} \notag \\
   +&i C^{I}\left(t_{2}-t_{1}\right) \hat{V}\left(t_{1}\right)^{\circ}\big]\big\}  \hat{\rho_{S}}(0)
  \end{align}
where $\hat{\rho}_{S}^{I}(t)$ is the reduced density matrix of system in the interaction
picture. Here, we consider the initial density matrix of the total system $\hat{\rho}_{tot}(0)=\hat{\rho}_{S}(0) \otimes \rho_{B}$, with $\hat{\rho}_{B}=\exp \left(-\beta H_{B}\right) / Z_{B}$
is the initial state of the bath. The superoperators $\hat{V}^{\times}$ and $\hat{V}^{\circ}$ correspond to commute and anti-commute operations, respectively.
At zero temperature, with the cavity initially in a vacuum state, the bath correlation function from Eq.~(\ref{Eq:spectrum}) can be decomposed into real and imaginary parts:
\begin{align}\label{Eq:appendix2}
  C^{R}(t_2 -t_1 )  = \sum_{k  = 1}^{2} \frac{\lambda}{2} e^{-\nu_{k} (t_2 -t_1)} \notag \\ C^{I}(t_2 -t_1)  = \sum_{k  = 1}^{2}(-1)^{k} \frac{\lambda}{2 i} e^{-\nu_{k}  (t_2 -t_1)}
  \end{align}
  where $\nu_{k}=\gamma+(-1)^{k} i \omega_{0}$ and $\omega_0$ is the characteristic frequency of the system.

  After inserting the correlation function Eq.~(\ref{Eq:appendix2}) into Eq.~(\ref{Eq:appendix1}) and repeatedly time deriving Eq.~(\ref{Eq:appendix1}), the HEOM for the spin-boson model is expressed as (More detail in Ref.~\cite{PhysRevA.85.062323})
  \begin{equation}\label{Eq:appendix3}
    \begin{aligned}
      \frac{d}{d t} \hat{\rho}_{l}(t)= & \left(-i \hat{H}_{s}^{\times}-\vec{l} \cdot \vec{v}\right) \hat{\rho}_{\bar{l}}(t)+\hat{\Phi} \sum_{p=1}^{2} \hat{\rho}_{\vec{l}+\vec{e}_{p}}(t) \\
      & +\sum_{p=1}^{2} l_{p} \hat{\Psi}_{p} \hat{\rho}_{\vec{l}-\vec{e}_{p}}(t)
      \end{aligned}
  \end{equation}
  with
\begin{equation}
  \hat{\Phi}=-i {V}^{\times},~~ \hat{\Psi}_{p}=-\frac{i}{2}\lambda\left[\hat{V}^{\times}+(-1)^{p} \hat{V}^{\circ}\right],
\end{equation}
here ${l}=(l_1,l_2)$ is a two-dimensional index, $\vec{v}=(\gamma-i\omega_0,\gamma+i\omega_0)$ is a two-dimensional vector, $\vec{e}_{1}=(1,0)$ and $\vec{e}_{2}=(0,1)$.
Besides, $\hat{\rho}_{\vec{l}=({0,0})}(0)=\hat{\rho}_{s}(0)$ is the reduced density matrix of system and $\hat{\rho}_{\vec{l}\neq({0,0})}$ is the auxiliary matrices with all elements are zero at initial moment.

\begin{figure}[htbp]
  \centering
  \includegraphics[scale=0.35]{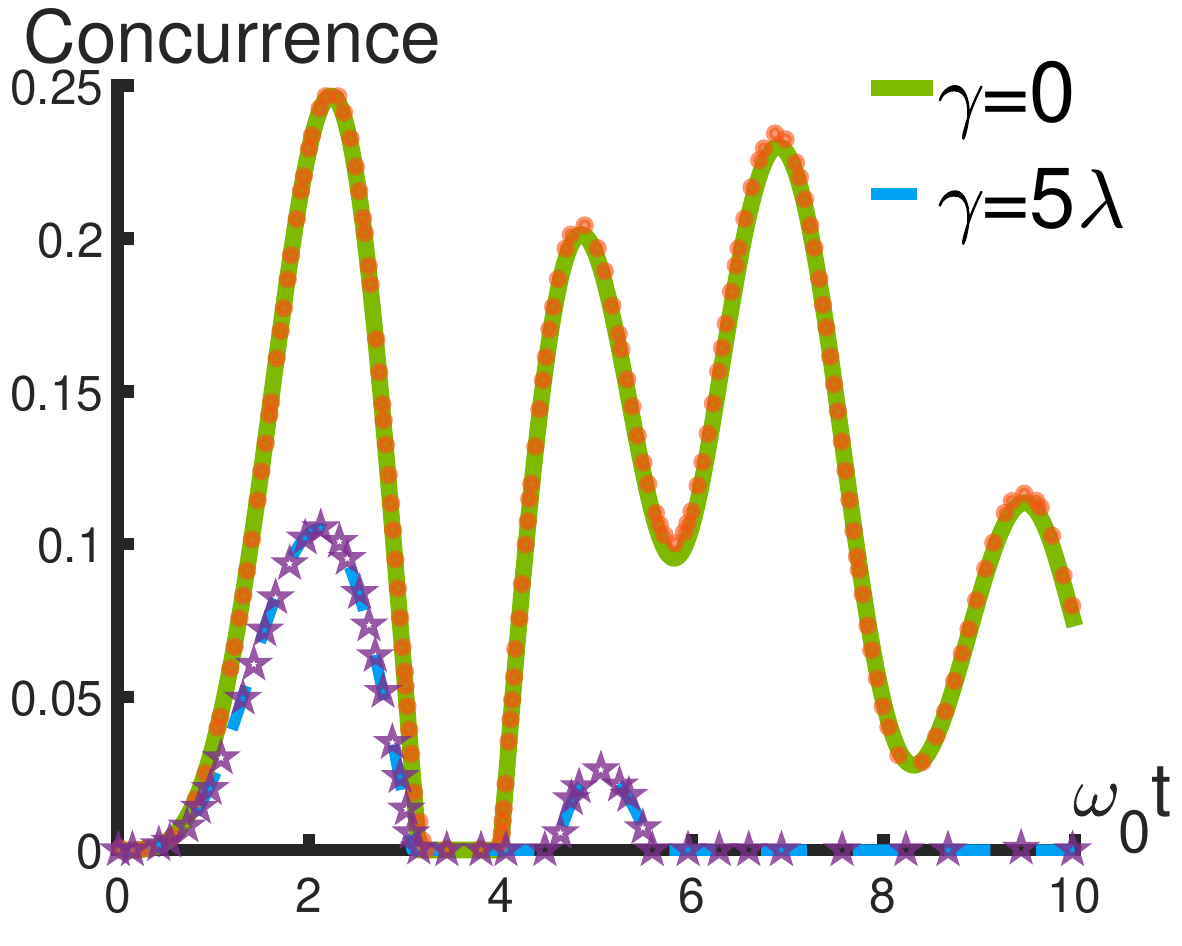}
  \caption{\label{fig:appendix1}The time evolution of the concurrence is calculated by our numerical HEOM program and the results from Ma Ji et al.~\cite{PhysRevA.85.062323} for various parameters: $\lambda=0$ (our numerical result: green solid line; Ma Ji et al.'s result: red points), $\lambda=5\lambda$ (our numerical result: blue solid line; Ma Ji et al.'s result: purple pentagrams); All other parameters are set to $\lambda=0.1\omega_0$ and $\omega_0=1$.}
  \end{figure}

The Eq.~(\ref{Eq:appendix3}) can be solved by standard numerical methods, such as the fourth-order Runge-kuta method, after the hierarchical equations are truncated for a sufficiently large integer L(e.g. $\hat{\rho}_{\vec{l}=(l_1,l_2)}(t)$ with $l_1+l_2>L$ will be dropped).
In addition, to demonstrate the validity of our HEOM program, we rigorously compared our results with those of Ma Ji et al.'s work~\cite{PhysRevA.85.062323}, which showed a high degree of agreement in Fig.~\ref{fig:appendix1}
\section{The exact dynamic equation of three qubits in a common bath under the RWA}\label{sec:secapp2}
In this section, we aim to derive the steps needed to obtain the exact time evolution equation for the reduced three-qubit system under the RWA.
To simplify the expression of the equation, we define the collective coupling constant $\alpha_{T}=\left(\alpha_{1}^{2}+\alpha_{2}^{2}+\alpha_{3}^{2}\right)^{1/2}$
and the relative strengths $r_{j}=\alpha_{j} / \alpha_{T}$.
By using the scheme proposed in Ref.~\cite{PhysRevA.79.032310}, we can obtain three integrodifferential equations for amplitudes $c_j(t)$ (j=1,2,3) as follow:
\begin{equation}\label{Eq:B1}
  \begin{aligned}
    \dot{c}_{1}(t)= & -\int_{0}^{t} d t_{1}\left[\alpha_{1}^{2} c_{1}\left(t_{1}\right)+\alpha_{1} \alpha_{2} c_{2}\left(t_{1}\right)+\alpha_{1} \alpha_{3} c_{3}\left(t_{1}\right) \right] \\
    & \times f\left(t-t_{1}\right) , \\
    \dot{c}_{2}(t)= & -\int_{0}^{t} d t_{1}\left[\alpha_{1} \alpha_{2} c_{1}\left(t_{1}\right) +\alpha_{2}^{2} c_{2}\left(t_{1}\right)+\alpha_{2} \alpha_{3} c_{3}\left(t_{1}\right) \right] \\
    & \times f\left(t-t_{1}\right) ,\\
    \dot{c}_{3}(t)= & -\int_{0}^{t} d t_{1}\left[\alpha_{1} \alpha_{3} c_{1}\left(t_{1}\right) +\alpha_{2}\alpha_{3} c_{2}\left(t_{1}\right) + \alpha_{3}^2 c_{3}\left(t_{1}\right) \right] \\
    & \times f\left(t-t_{1}\right) ,
    \end{aligned}
\end{equation}
where the correlation function $f\left(t-t_{1}\right)$ is defined as the Fourier transform of the spectral density $J(\omega)$ and we can derive the quantities $c_j(t)$ by performing the Laplace transform of Eqs.~(\ref{Eq:B1})
yield
\begin{equation}
  \begin{aligned}
    s \widetilde{c}_{1}(s)-c_{1}(0)= & -\left[\alpha_{1}^{2} \widetilde{c}_{1}(s)+\alpha_{1} \alpha_{2} \widetilde{c}_{2}\left(s\right)+\alpha_{1} \alpha_{3} \widetilde{c}_{3}\left(s\right)\right] \\
    & \times \widetilde{f}\left(s\right), \\
    s \widetilde{c}_{2}(s)-c_{2}(0)= & -\left[\alpha_{1} \alpha_{2} \widetilde{c}_{1}\left(s\right)+\alpha_{2}^{2} \widetilde{c}_{2}(s)+\alpha_{2} \alpha_{3} \widetilde{c}_{3}\left(s\right)\right] \\
    & \times \widetilde{f}\left(s\right), \\
    s \widetilde{c}_{3}(s)-c_{3}(0)= & -\left[\alpha_{1} \alpha_{3} \widetilde{c}_{1}\left(s\right)+\alpha_{2}\alpha_{3} \widetilde{c}_{2}(s)+ \alpha_{3}^2 \widetilde{c}_{3}\left(s\right)\right] \\
    & \times \widetilde{f}\left(s\right).
    \end{aligned}
  \end{equation}
  In order to simplify the solution of the equation,  we define $\left|\psi_{+}\right\rangle=r_{1}|1\rangle_{1}|0\rangle_{2}|0\rangle_{3}+r_{2}|0\rangle_{1}|1\rangle_{2}|0\rangle_{3}+r_3|0\rangle_{1}|0\rangle_{2}|1\rangle_{3}$
  with a structure analogous to the solution of the case presented in Ref.~\cite{PhysRevA.79.032310} and $\mathcal{E}(t) = \left\langle\psi_{+} \mid \psi_{+}(t)\right\rangle$ is given by 
  \begin{align}
    \mathcal{E}(t) & = e^{-\lambda t / 2}\left[\cosh (\Omega t / 2)+\frac{\lambda}{\Omega} \sinh (\Omega t / 2)\right],
    \end{align}
    Here $\Omega=\sqrt{\lambda^{2}-4 \mathcal{R}^{2}}$. Based on the expression of $\mathcal{E}(t)$, we can get the exact solution of the probability amplitude $c_j$ as
\begin{equation}
\begin{aligned}
  c_{1}(t) & = \left[r_{2}^{2}+r_{3}^{2}+r_{1}^{2} \mathcal{E}(t)\right] c_{1}(0)-r_{1} r_{2}[1-\mathcal{E}(t)] c_{2}(0)\\ &-r_{1} r_{3}[1-\mathcal{E}(t)] c_{3}(0), \\
  c_{2}(t) & =\left[r_{1}^{2}+r^2_3+r_{2}^{2} \mathcal{E}(t)\right] c_{2}(0) -r_{1} r_{2}[1-\mathcal{E}(t)] c_{1}(0) \\ &-r_{2} r_{3}[1-\mathcal{E}(t)] c_{3}(0), \\
  c_{3}(t) & =  \left[r_{1}^{2}+r_{2}^{2}+r_{3}^{2} \mathcal{E}(t)\right] c_{3}(0)-r_{1} r_{3}[1-\mathcal{E}(t)] c_{1}(0)\\ & -r_{2} r_{3}[1-\mathcal{E}(t)] c_{2}(0).
  \end{aligned}
\end{equation}
Thus, the evolution of the entanglement dynamics in the system is influenced by the coupling strength $r_j$, $\mathcal{E}(t)$ and the initial state $c_j(0)$.

\bibliography{manuscript.bib}
\end{document}